\documentclass[final,5p,twocolumn]{elsarticle}

\usepackage{graphicx}

\usepackage{tabularx}
\usepackage{bbold}

\usepackage{subfig}

\usepackage{lineno,hyperref}
\modulolinenumbers[5]

\journal{Computer Physics Communications}

\biboptions{numbers,sort&compress}

\begin{document}

\begin{frontmatter}

\title{Pole-fitting for complex functions: enhancing standard techniques by artificial-neural-network classifiers and regressors}

\author[oeaw,unigraz]{Siegfried Kaidisch}

\author[zeiss]{Thomas U. Hilger}

\author[oeaw,unigraz]{Andreas Krassnigg\corref{mycorrespondingauthor}}
\cortext[mycorrespondingauthor]{Corresponding author}
\ead{andreas.krassnigg@uni-graz.at}

\author[oeaw]{Wolfgang Lucha}

\address[oeaw]{Institute for High Energy Physics, Austrian Academy of Sciences, Nikolsdorfergasse 18, A-1050 Vienna, Austria}
\address[unigraz]{University of Graz, Institute of Physics, NAWI Graz, A-8010 Graz, Austria}
\address[zeiss]{Semiconductor Mask Solutions, Carl Zeiss SMT GmbH, Carl-Zeiss-Promenade 10, 07745 Jena, Germany}

\begin{abstract}
Motivated by a use case in theoretical hadron physics, we revisit an application of a pole-sum fit to dressing functions of a confined quark propagator.
More precisely, we investigate approaches to determine the number and positions of the singularities closest to the origin for a function that is only known numerically on a specific finite grid of values on the positive real axis. 
For this problem, we compare the efficiency of standard techniques, like the Levenberg-Marquardt algorithm, to a pure artificial-neural-network approach as well as a combination of these two. 
This combination is more efficient than any of the two techniques separately. 
Such an approach is generalizable to similar situations, where the positions of poles of a function in a complex variable must be quickly and reliably estimated from real-axis information alone.
\end{abstract}


\end{frontmatter}


\section{Introduction}\label{introduction}

In the standard model of particle physics, one type of fundamental interaction is referred to as the strong interaction among and within its subjects called hadrons.
In the setting of quantum chromodynamics (QCD), it explains strong-interaction phenomena in terms of quarks, gluons, and their interactions, respectively.
While QCD is accessible by perturbation theory in the high-energy domain \cite{Gross:1973ju,Politzer:1974fr}, nonperturbative methods are required for the study of bound states.

One nonperturbative avenue is to use the set of Dyson-Schwinger equations (DSEs)  \cite{Dyson:1949ha,Schwinger:1951ex,Schwinger:1951hq}, an infinite tower of coupled and, in general, nonlinear integral equations.
Bound states or other low-energy phenomena are treated in this approach via covariant equations of the Bethe-Salpeter-equation (BSE) type \cite{Bethe:1951bs,Salpeter:1951sz}, or analogous equations for a larger number of constituents in the bound state.

In practice, such a set of equations is usually truncated in order to render it tractable by numerical methods. 
In essence, this results in a finite set of coupled integral equations, where solutions of one equation appear as input to other equations in the set.
A particular difficulty in this numerical endeavour is the need to solve the set self-consistently across equations, and for complex values of, at least, some variables.

The latter point is delicate insofar as one may have to perform an analytic continuation of one equation's solution from the positive real axis into the complex plane relying on numerical methods.
The crux with such a step is connected to the possible appearance of singularities in the relevant complex domains.
While this isn't problematic in principle and, in fact, is to be expected in the presence of relevant physical scales such as particle masses, it gets in the way of iterative solution procedures, series expansions, or other possible ways towards a solution of the problem at hand.

Herein, we have revisited a concrete use case in hadron physics which exhibits this kind of situation and compared the following approaches:
\begin{itemize} 
  \item standard ways of fitting the real-axis solution to a sum of poles that is then used to represent the solution in the complex plane,
  \item an artificial neural network (ANN) for classification of pole configuration and subsequent regression for pole positions, and
  \item a combination of the two.
\end{itemize}

As it turns out, the standard approach and the ANN prediction can inform each other and increase the method's efficiency and success rate.
This technique can be immediately generalized within the DSE framework, but there is no reason why it could not be helpful in many similar cases of numerical problems across different fields in physics and natural sciences in general.

\section{Concrete use case and setup}\label{usecase}

\paragraph{DSE for the quark propagator}
The concrete use case and the origin of our study is the DSE for the dressed quark propagator in QCD.
This equation relates several Green functions, namely the bare ($S_0$) and dressed ($S$) quark propagators, the dressed gluon propagator $D$, and the quark-gluon vertex in its bare ($\gamma$) and dressed ($\Gamma$) forms.
Pictorially, one has
\begin{equation}
  \parbox[c]{0.85\columnwidth}{\includegraphics[width=0.82\columnwidth]{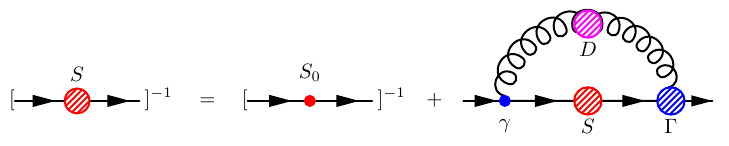}}
\end{equation}
Written as an equation of momenta flowing through the above diagrams, neglecting issues like regularization or renormalization as well as internal quantum numbers, it has the form
\begin{equation}\label{eq:dse}
  S^{-1}(p)  =  S_0^{-1}(p) +  \int_q  D^{\mu\nu}(k)\, \gamma^\mu \,S(q)\,  \Gamma^\nu(q,p)\,  .
\end{equation}
Herein, $S(p)$ is the renormalized dressed quark propagator, $S_0(p)$ is its bare counterpart.
$D^{\mu\nu}(k)$ is the renormalized dressed gluon propagator.
The renormalized dressed and bare quark-gluon vertices are denoted by $\Gamma^\nu(q,p)$ and $ \gamma^\mu$, respectively.
The momenta $p$ and $k$ are the quark and gluon momenta, respectively, and $q=p-k$ is an integration momentum with $\int_q = \int d^4q/(2\pi)^4$ representing the corresponding four-momentum integration.

\paragraph{Quark-propagator dressing functions}
For many studies, in particular of mesons and baryons as bound states in QCD, Eq.~(\ref{eq:dse}) is an essential ingredient.
More concretely, its solution, the dressed quark propagator, appears inside integrals in covariant bound-state equations, whose numerical solution is a standard approach to the problem of studying relativistic bound states \cite{Blank:2010bp}.
The covariant structure of the dressed quark propagator is that of a fermion, namely (see, e.\,g., \cite{Sanchis-Alepuz:2017jjd})
\begin{equation}\label{eq:dressedpropagator}
S(p) = -i\; \gamma\cdot p \;\sigma_{v}(p^{2})+\mathbb{1} \sigma_{s}(p^{2}),
\end{equation}
where the two dressing functions are usually written in relation to the dressing functions $A$ and $B$ of the corresponding inverse quark propagator as
\begin{equation}\label{eq:sigmav}
\sigma_{v}(p^{2})=\frac{A(p^{2})}{p^{2} A^{2}(p^{2})+B^{2}(p^{2})}
\end{equation}
and
\begin{equation}\label{eq:sigmas}
\sigma_{s}(p^{2})=\frac{B(p^{2})}{p^{2} A^{2}(p^{2})+B^{2}(p^{2})}\;.
\end{equation}
These forms show the dependence of the propagator on Dirac-scalar functions ($A$, $B$, $\sigma_{v}$, and $\sigma_{s}$), which effectively characterize a numerical solution, and for which numerical coupled equations emerge via projections on the relevant Dirac structures.
As a result, one basically deals with solving coupled equations of a set of dressing functions for each of the propagators, kernels, and amplitudes involved in the original set of coupled covariant equations.

\paragraph{Role of dressing functions in bound-state and other studies}
The covariant structure of these equations makes it straightforward to set total momenta of the system under consideration and to establish a set of consistent four-momenta throughout all equations.
In doing this, these four-momenta and their relations to each other determine all arguments of the dressing functions in all parts of the equations.
In particular, one has to deal with complex variables in some cases due to the four-dimensional nature and the metric properties of space-time, whether in its Minkowski \cite{Sauli:2018gos} or Euclidean \cite{Eichmann:2009zx} variants, the latter of which is used herein.

In essence, the freedom associated with choosing loop-integration momenta in quantum field theory can be used to keep the arguments of some dressing functions real, but it is very common to end up with several complex arguments in dressing functions, in particular of the quark propagators \cite{Mader:2011zf}, Bethe-Salpeter amplitudes \cite{Bhagwat:2006pu}, or dressed vertices \cite{Williams:2015cvx} involved in the relevant diagrams.
In particular, instances of the dressed quark propagator embedded in bound-state studies require the continuation of real-axis solutions for $\sigma_{v}(p^2)$ and $\sigma_{s}(p^2)$ into the complex plane of quark momentum squared $p^2$ .

\paragraph{Quark-propagator dressing functions of complex arguments}
For practical purposes, such a continuation in a momentum-squared variable does not necessarily need to be rigorous in a mathematical sense.
On the contrary, various approximation methods can be used to get an estimate for function values away from the real momentum-squared axis, on which solutions are normally known.
As the most simple version of this, the function value on the real axis for each or one specific real part can be defined to serve for all or surrounding corresponding imaginary parts of momentum squared in what is usually referred to as a \emph{real-axis approximation}, e.g., \cite{Souchlas:2010zz}.

Another version of the \emph{real-axis approximation} is to use series expansions of the dressing functions at several points on the real axis to compute function values of complex arguments close to these points. 
Such efforts have already been successfully applied several years ago \cite{Jain:1993qh}.
In a BSE setting, this kind of \emph{Ansatz} is often combined with Bethe-Salpeter amplitudes expanded in a series of Chebyshev polynomials and their moments \cite{Munczek:1991jb,Maris:1999nt,Krassnigg:2003dr}, where complex arguments can be fed directly into the relevant terms of the polynomial sum.

However, the most reliable efforts in this direction come from direct numerical continuations of the quark-propagator dressing functions in finite regions of complex momentum squared \cite{Fischer:2005en,Krassnigg:2008gd,Eichmann:2009zx,Windisch:2019byg}.

It should also be mentioned here that reducing the set of coupled integral equations to a set of algebraic equations can be achieved via a very simple interaction structure \cite{Munczek:1983dx}.
In such a setup, complex arguments can be handled more easily \cite{Bhagwat:2004hn,Gomez-Rocha:2016cji}.

\paragraph{Singularities in quark-propagator dressing functions}
With all these numerical methods to arrive at complex values for arguments of the dressing functions of the quark propagator, we have a rather solid understanding of what these functions look like in the complex domain, and how reliable the various kinds of approximations are in practical calculations.
However, numerical results are not always easy to handle in the BSE.
In particular, if singularities appear to be present in the complex sampling domain of the BSE kernel, one must resort to some alternative to a purely numerical treatment \cite{Bhagwat:2002tx}.

In Ref.~\cite{Bhagwat:2002tx}, the authors constructed a version of the BSE kernel where the momentum integrals could be performed using Feynman-integration techniques for the singularities present in the complex sampling domain. 
To make this work, they had to assume a particular structure of these singularities.
Their \emph{Ansatz} in this case was a sum of three pole pairs with complex conjugate values for the pole positions in each pair.
While their parameterization was based on a sum of several confined fermion propagators with a mass scale each and looks a bit different, one can use the following simple form to achieve this:
\begin{equation}\label{eq:complexpolesum}
f\left(z\right)=\sum_{n=1}^{N_{c}}\left\{\frac{c_{n} }{z-v_{n}}+\frac{c_{n}^{*}}{z-v_{n}^{*}}\right\} \;,
\end{equation}
where
\begin{equation}
z, c_n, v_n \in \mathbb{C}.
\end{equation}
With an \emph{Ansatz} like this, the remaining problem is to find the values for the parameters $c_n, v_n$. 
In principle, the number of complex pole pairs $N_{c}$ is a free parameter as well, and we will discuss this below.
For the moment, it is sufficient to mention that the authors of Ref.~\cite{Bhagwat:2002tx} found $N_{c}<3$ to be inadequate for a satisfactory fit of Eq.~(\ref{eq:complexpolesum}) to their real-axis solutions for the quark DSE.
As a consequence, they settled on a sum of three complex-conjugate pole pairs in order to describe the complex behavior of the quark dressing functions. 
They then fitted the remaining free parameters to represent the real-axis solution.

\begin{figure}[ht]
\centering
\includegraphics[trim=170 0 90 20bp,clip,width=1.0\linewidth,keepaspectratio]{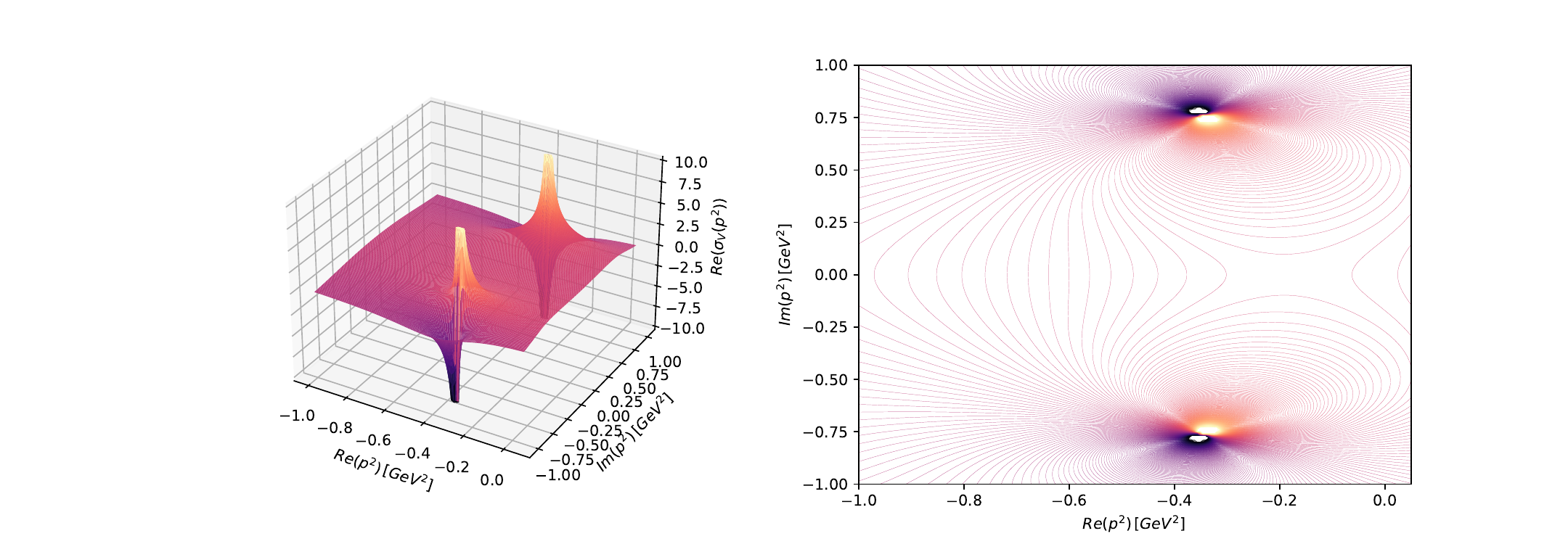}
\caption{
Illustration of the actual pole structure in a practical example calculation, where the dominant (closest to the origin) singularities are a complex-conjugate pole pair. }
\label{fig:propagatorexample_1}
\end{figure}

\begin{figure}[ht]
\centering
\includegraphics[trim=170 0 90 20bp,clip,width=1.0\linewidth,keepaspectratio]{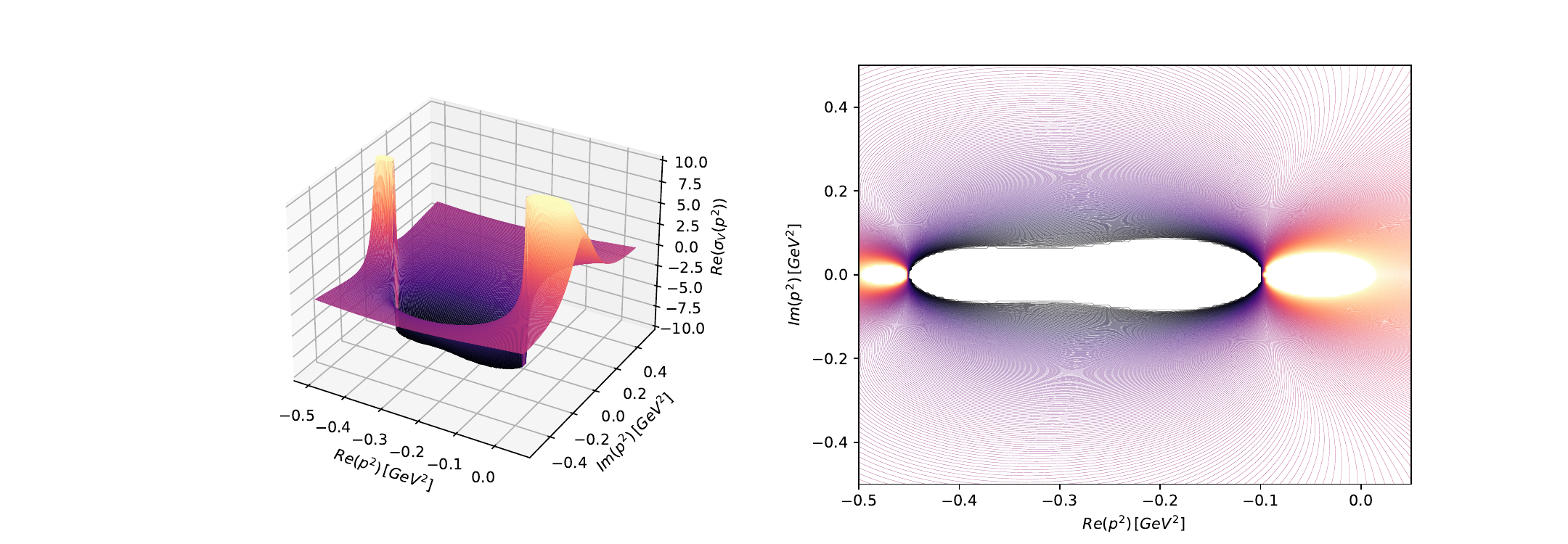}
\caption{
Illustration of the actual pole structure in a practical example calculation, where the dominant (closest to the origin) singularities lie on the real axis. }
\label{fig:propagatorexample_2}
\end{figure}

\paragraph{Example quark-propagator dressing functions}
Before we go into the details of how to deal with the problem of finding the configuration and positions of poles in a sum like the one in Eq.~(\ref{eq:complexpolesum}), a reality check is in order. 
In a number of studies of bound states, the actual analytic structure of the quark-propagator dressing functions has been mapped out numerically. 
However, these computations are tedious and do not provide much benefit in terms of arriving at a definite answer for a parameterization such as Eq.~(\ref{eq:complexpolesum}).
On the other hand, one can plot the dressing functions for complex arguments by a brute-force numerical integration, albeit forgoing mathematical rigor to a certain extent. 
As a consequence, the plots in Figs.~\ref{fig:propagatorexample_1} and \ref{fig:propagatorexample_2}  are meant mainly as illustrations of the problem at hand.

\begin{figure}[ht]
\centering
\includegraphics[trim=170 0 90 20bp,clip,width=1.0\linewidth,keepaspectratio]{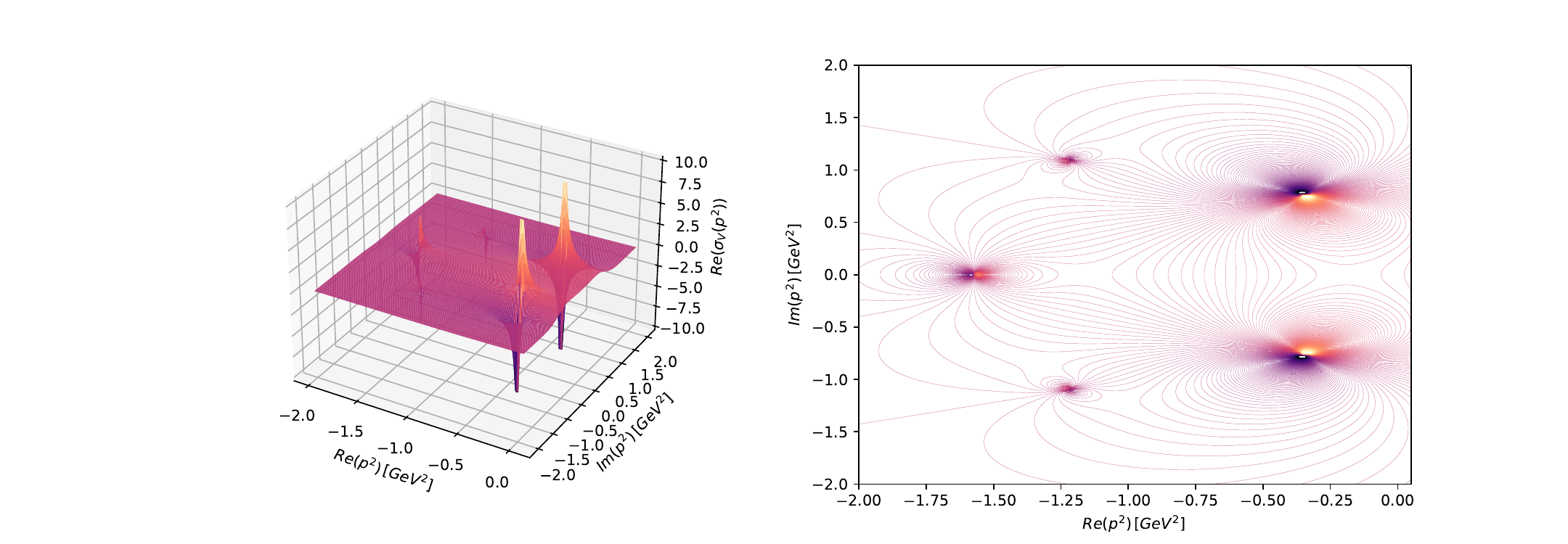}
\caption{
Illustration of the actual pole structure in a practical example calculation with both real and complex-conjugate singularity positions. }
\label{fig:propagatorexample_3}
\end{figure}

Our first example, shown in Fig.~\ref{fig:propagatorexample_1}, seems to conform to the complex-pole-sum assumption. 
The figure has two parts with a 3D-plot on the left and a contour plot on the right.
The real part of $\sigma_{v}(p^2)$ is shown on the complex-$p^2$ domain where 
\begin{equation}
  \Re (p^2) \in [-1,0.05], \qquad \Im (p^2) \in [-1,1]\;.
\end{equation}  
Another example (resulting from different model parameters) is shown in Fig.~\ref{fig:propagatorexample_2}, where
\begin{equation}
  \Re (p^2) \in [-0.5,0.05], \qquad  \Im (p^2) \in [-0.5,0.5]\;,
\end{equation}  
and a different situation emerges.
In particular, in this case the dominant singularities lie on the real axis.
While this is not a problem per se in the calculation or even the bound-state treatment, it does necessitate an amendment to the \emph{Ansatz} given in Eq.~(\ref{eq:complexpolesum}).
This necessity becomes even more evident when our first example is inspected on a larger complex domain, which is shown in Fig.~\ref{fig:propagatorexample_3}, where one can see a mix of both complex-conjugate and real positions of the apparent singularities.

On a small, but important side note: real-world examples like the ones above show upon investigation that the singularities in the quark propagator appear via zeros in the common denominator in Eqs.~(\ref{eq:sigmav}) and (\ref{eq:sigmas}).
The dressing functions $A$ and $B$, on the other hand, are regular on the domain of consideration.
Thus, the same singularities are found in both the real and imaginary parts of both $\sigma_{v}$ and $\sigma_{s}$.

Overall, the necessary consequence is to include poles with real positions in the sum of Eq.~(\ref{eq:complexpolesum}), which is achieved by Eq.~(\ref{eq:polesum}) below.
One may question the reliability of such an \emph{Ansatz} for properly mimicking a function's behavior on the complex domain, when fitting the free parameters to a numerical expression known only on the real-$p^2$ axis.
In particular, at this point the question arises how accurate or reliable such a fit can be if the \emph{Ansatz} does not contain the type of singularity actually present in the numerical solution.
In other words, can one fit a complex-conjugate pole sum by a sum of real poles, or vice versa, and obtain a real-axis agreement that seems satisfactory, as presented in Ref.~\cite{Bhagwat:2002tx}?

In the following, we approach an answer to this question and others by detailing the necessary steps.
We discuss standard fitting techniques, confront them with the application of artificial neural networks, and finally combine these two approaches in an attempt to obtain an efficient and reliable method for finding, in particular, the dominant singular contribution in a problem such as the one described here.

\section{Standard approaches to the problem}\label{standard}

\paragraph{General considerations}
In an attempt to fit the singularity structure of a function to numerical data only (i.e., without much analytical knowledge), there are two steps:

Firstly, one needs information about the number and kind of singularities within the relevant domain.
For example, it is assumed herein (for the sake of simplicity, but also based on observations from visualizations like the ones above) that we are dealing with simple poles and not any kind of singularity of higher or non-integer order.

Secondly, once the number and kind of singularities are known or assumed, standard numerical fitting techniques can be applied to obtain estimates of both the positions and the coefficients of each singularity in the sum.
Note that these coefficients are complex conjugate numbers for complex conjugate pole pairs and real for real poles as a result of the physical restrictions on function values on the positive real axis.
Note also that, while scenarios with cuts along parts of the negative real axis are conceivable, we ignore such cases herein.

As a result, we work with a sum of simple poles and pole pairs.
In a straightforward approach for obtaining the free parameters of the pole sum, i.e., the poles' positions and coefficients, first the numbers of both real and complex poles are fixed to reasonable values.
In a second step, conventional curve-fitting methods can then be used to fit the pole sum's parameters to the data.

The general form for such a fit used herein is
\begin{equation}\label{eq:polesum}
f\left(z\right)=\sum_{n=1}^{N_{c}}\left\{\frac{c_{n} }{z-v_{n}}+\frac{c_{n}^{*}}{z-v_{n}^{*}}\right\} + \sum_{n=1}^{N_{r}}\left\{\frac{d_{n} }{z-w_{n}}\right\},
\end{equation}
where
\begin{equation}
z, c_n, v_n \in \mathbb{C},~d_n, w_n \in \mathbb{R}.
\end{equation}

\paragraph{Conventional fitting methods}
For comparison and evaluation, we at first tested three different iterative algorithms which are easily available from the SciPy Python-package \cite{2020SciPy-NMeth}:

\begin{itemize} 
  \item Levenberg-Marquardt algorithm (LM) \cite{Levenberg:1944lm,Marquardt:1963lm}: this algorithm showed the highest accuracy of the three listed here, but may yield unreliable results for our setup (see below).
  \item Trust region reflective algorithm (TRF) (see, e.g., \cite{Yuan:2015trf} for a recent review): for our problem, this algorithm has good accuracy and it always converges.
  \item Dogleg algorithm with rectangular trust regions (DOGBOX) \cite{Powell:1970dogleg,VoglisLagaris:2014dogbox}: for our problem, this algorithm has the lowest accuracy in comparison, but it always converges.
\end{itemize}

For the problem presented in this manuscript, (bounded) TRF proved to be the most suitable method out of these three, since the algorithm always converged (in contrast to unbounded LM, when applied to real-world data, see below) and outperformed DOGBOX accuracy-wise. 
However, a further discussion of the boundary conditions of the fitting methods is in order.
In fact, we know from the physical use case that the appearance of singularities close to the origin is to be expected.
In particular, the appearance of mass scales from a few MeV up to a few GeV in dressed-quark propagators are physically reasonable, although quarks are confined in QCD, i.e., one does not expect the usual simple mass pole found in the free particle propagator in Minkowski space.
Actually, Ref.~\cite{Bhagwat:2002tx} argues how complex conjugate pole pairs make a case for quark confinement inside a hadron.

In summary, we assumed a general data structure such that there should be singularities within a certain finite and quite definite area of the complex quark-momentum squared, which is close to the origin and constrained to the half plane of negative real parts of the $p^2$ variable.
As a result, we can put an emphasis on bounded fitting methods, both in the hopes for (faster) convergence and for better accuracy. 

\paragraph{Bounded vs. unbounded fitting methods}
As can be seen in more detail from Tab.~\ref{tabcompareregression} further below, the unbounded LM algorithm massively outperforms both the bounded TRF and the bounded DOGBOX algorithms on synthetic data. 
For now, however, we want to draw attention to another issue.
In addition to other problems with the unbounded LM algorithm even on synthetic training data, which are explained below in Sec.~\ref{anns}, the unbounded LM algorithm never converged when applied to real-world data.

Since bounded execution of an LM fit is not implemented in SciPy directly, an additional Python library, LMFIT \cite{Newvilleetal:2014lmfit}, was used. 
LMFIT offers a wide variety of (bounded) fitting methods, and so we took the opportunity to also test the following for the sake of a wider comparison:

\begin{itemize} 
  \item Bounded version of the LM algorithm (LEASTSQ)
  \item Basin-hopping algorithm (BASINHOPPING) \cite{WalesDoye:1997bh}
  \item Broyden-Fletcher-Goldfarb-Shanno Hessian update strategy (BFGS) \cite{Broyden:1970bfgs,Fletcher:1970bfgs,Goldfarb:1970bfgs,Shanno:1970bfgs}
  \item Adaptive Memory Programming for Global Optimization (AMPGO) \cite{Lasdonetal:2010ampgo}
  \item Powell algorithm (POWELL) \cite{Powell:1964pow}
  \item Truncated Newton algorithm (TNC) \cite{Nash:1984tmc,Nash:1985tmc}
  \item Dual Annealing optimization (DUAL\_ANNEALING) \cite{Tsallis:1996dual,Xiang:1997dual,Xiang:2000dual}
  \item Conjugate-Gradient algorithm (CG) \cite{Hestenes:1952cg}
  \item L-BFGS-B algorithm (LBFGSB) \cite{Byrdetal:1995lbfgs,Zhu:1997lbfgs}
  \item Nelder-Mead algorithm (NELDER) \cite{Nelder:1965nm}
  \item Sequential Linear Squares Programming (SLSQP) \cite{Nocedal:2006no}
  \item Constrained Optimization BY Linear Approximation (COBYLA) \cite{Powell:2007cobyla}
  \item Differential evolution method (DIFFERENTIAL\_EVOLUTION) \cite{Storn:1996de,Storn:1997de}
\end{itemize}

Detailed results for the fit accuracy of all these compared to our ANN approach are shown below.
For the moment, we would like to mention that out of all methods tried, the LM algorithm was the only one which yielded better results (lower errors of the fitted parameters) when used without boundaries.
All other methods (TRF, DOGBOX and all other LMFIT methods listed above) showed better results when used in a bounded way.
The performance of all tested bounded methods on synthetic training data is shown below in Fig. \ref{fig:comparison}, while the performance of unbounded LM can be taken from Tab. \ref{tabcompareregression}.

\paragraph{Problems of conventional methods}

The most apparent problem in such an approach is the need to somehow choose and fix the number of poles manually. 
In particular, there is no easy way of deducing appropriate values for $N_{c}$ and $N_{r}$ directly from the data, on which the fits are to be performed.
It is unsatisfactory to simply rely on convergence and add more and more poles, i.e., increase  $N_{c}$ and/or $N_{r}$ until the result does not change by large amounts any more or until the numerical error between fit and the data is smaller than some margin, in particular, because such a method becomes numerically unstable when the number of free parameters gets too large.
In Sections \ref{anns} and \ref{combination}, we present ANN classifiers that aim at solving this problem by predicting the layout of the dominant (i.e., closest to the origin) singularities for a given numerical input set of function values on the positive real axis.

\begin{figure}[ht]
\centering
\includegraphics[width=1.0\linewidth,keepaspectratio]{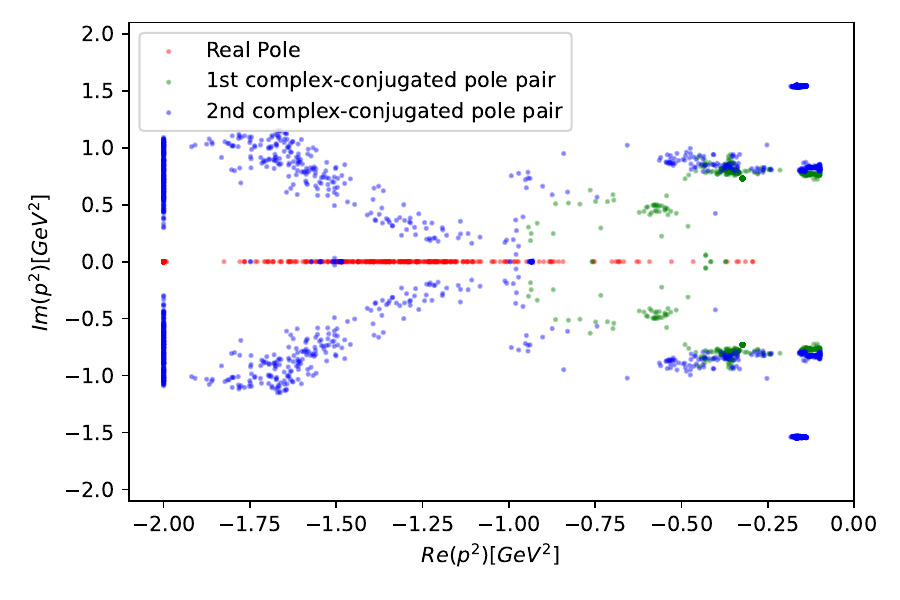}
\caption{
Using different initial values each time, a pole sum consisting of one real pole and two complex-conjugated pole pairs was fitted to the same real-world data 1000 times using the bounded TRF algorithm. 
This scatter plot shows the pole positions (color-coded with respect to the corresponding pole) resulting from these fits, scattered across the complex plane.
The spread is a result of the variety of initial conditions for the fit parameters. See text for a discussion.}
\label{fig:scatterplot}
\end{figure}

Another, less apparent problem of this approach is the strong dependence of the resulting pole parameters on their initial values as they are used by the fitting algorithm.
In practice, running the same fitting algorithm several times on the same data, the results can differ due to randomly selected initial guesses for the parameters.
In particular, even if the correct number of real and complex poles is known, different initial guesses can lead to vastly different converged values for the pole parameters.
To illustrate this, we have plotted the resulting parameters of such a group of runs in Fig.~\ref{fig:scatterplot}.

As a consequence, conventional fitting should always be combined with an analysis of the dependence of the results on initial parameter values.
However, even if such an analysis were performed, care would have to be taken to (in our case) correctly conclude what the best candidates for the pole positions would be.
At least, distribution plots such as Fig.~\ref{fig:scatterplot} appear to be somewhat inconclusive for a precision analysis.
It should be noted that the bands on the left border of the figure result from the fact that the fit results were constrained, and not allowed to go beyond the boundary of $\Re (p^2) = -2\mbox{ GeV}^2$.

\section{Using artificial neural networks}\label{anns}

\paragraph{Using ANNs to solve the problem}
In order to fit the data to a pole sum with an a priori unknown number of poles, a combination of an ANN classifier and an ANN regressor can be used.
The classifier is used to determine the number of real poles $N_r$ and complex-conjugated pole-pairs $N_c$ corresponding to the data. 
For the purpose of this paper, the total number of poles was restricted by setting $N_r + N_c \leq 3$ (this upper limit may be set to a higher value, depending on the application).
This results in nine possible combinations of $N_r$ and $N_c$, each of which is then identified as a separate class of the classification problem.

After $N_r$ and $N_c$ have been determined by the classifier, a neural network regressor (or a conventional fitting method) can be used to determine the pole parameters corresponding to the data. 
Note that for each class, that is, for each combination of $N_r$ and $N_c$, a separate regressor must be trained, since different summation limits $N_r$ and $N_c$ in a pole sum lead to different regression problems for finding the pole parameters.

\paragraph{Synthetic data}
The performance of the approaches presented in this paper can easily be calculated and compared to each other by applying them to synthetic data, whose exact parameters are known beforehand. 
We used such synthetic data for training, validating, and testing our various ANN classifiers and regressors as well as the combined approach described further below.

Concretely, one can calculate synthetic curve data by manually selecting values for $N_r$ and $N_c$, as well as randomly selecting the pole positions and coefficients, entering these values into the pole sum, and thus evaluating the expression in Eq.~(\ref{eq:polesum}) on a pre-defined momentum-squared grid.
Each piece in such a set of curve data is a suitable input for both the neural networks as well as the conventional fitting methods.
Conveniently, these data are very cheap to generate in large quantities.
As a result, enough training data can easily be generated and tailored to the needs of the specific problem at hand.

During ANN training (and validation/testing of ANNs and conventional methods), the outputs ($N_r$, $N_c$, fitted/predicted parameters of the pole sum) of each prediction/fit can be compared directly to the values that were originally used to create the curve, thus revealing the performance of the respective method.
It should, however, be noted that training a neural network on synthetic data and then applying it to real-world data only constitutes a sound approach if the synthetic data is {\it similar enough} to the real-world data (see Sec. \ref{application}).
More details on the concept of synthetic data can be found, e.g., in \cite{Leetal:2017sd}.

In more detail, the data creation can be performed in the following manner:

\begin{enumerate}
\item Set boundaries for $N_r$, $N_c$, and the pole positions and coefficients:
in this section, as well as in Section \ref{combination}, the following boundaries were used (unless stated differently):

\begin{itemize}
\item $N_r + N_c \leq 3$
\item $ \Re\left(v_n\right) \in \left[-10,-0.1\right], \Im\left(v_n\right) \in \left[0,10\right]$
\item $w_n \in \left[-10,-0.1\right]$
\item $\Re\left(c_n\right) \in \left[-1,1\right], \Im\left(c_n\right) \in \left[-1,1\right]$
\item $d_n \in \left[-1,1\right]$
\end{itemize}

\item For each allowed combination of $N_r$ and $N_c$, generate $N_\mathrm{data}$ sets of pole parameters, with values chosen randomly within the parameter boundaries.
\item Enter each of these sets of parameters---together with the corresponding values for $N_r$ and $N_c$---into the pole sum to create a formula for obtaining a curve from a set of momentum-squared points.
\item Generate each data curve on the momentum-squared (positive) real-axis grid used (for specific details of this grid, we refer the interested reader to Appendix C of Ref.~\cite{Hilger:2017jti}).
\end{enumerate}

\paragraph{Using data of both $\sigma_{v}$ and $\sigma_{s}$ combined} 
As described in Sec.~\ref{usecase}, the same singularities are found in both the real and imaginary parts of both $\sigma_{v}$ and $\sigma_{s}$.
That is, we know \emph{a priori} that $\Re (\sigma_{v})$ and $\Re (\sigma_{s})$ share the same pole structure, i.e., the number and position of any real and/or complex poles described by the parameters $N_r$, $N_c$, $v_n$, $w_n$, and differ only in the pole coefficients ($c_n$, $d_n$).
For clarity, we note here that the imaginary parts $\Im (\sigma_{v})$ and $\Im (\sigma_{s})$ cannot be used, since the expression in Eq.~(\ref{eq:polesum}) is real on the real axis in our setup.

Using these two curves at the same time instead of using only a single data curve enhances the various methods discussed in this paper.  
For this reason, we used both curves in all methods presented rather than using only one of them. 
For example, for the conventional fitting methods, both curves are simultaneously used as inputs and the fitting methods return the common pole positions ($v_n$, $w_n$) as well as the individual pole coefficients ($c_{n,1}$, $d_{n,1}$, $c_{n,2}$, $d_{n,2}$) for each curve.

However, in order not to overcomplicate the description of the various methods and since this detail is not essential for the understanding of the presented ideas, it is omitted in the rest of this paper and we only talk about (individual) data curves or pole sums. 
The interested reader can find additional information on this kind of detail in the implementation of the described methods, see Sec. \ref{code}.

\paragraph{Classifier}
A simple feedforward ANN classifier (for an introduction, see, e.g., \cite{Svoziletal:1997ff}) that directly takes pole-curve values on our momentum-squared grid points as its input and returns $N_r$ and $N_c$ was trained and tested on synthetic data as described above.
This classifier, as well as all other neural networks in this paper, were trained using the Adam optimization algorithm \cite{Kingma:2015adam}. 

In terms of the network architecture, we used a standard setup with fully connected (linear) layers. 
In addition to the input and output layers, we tested several numbers of hidden layers with rectified linear unit (ReLU) \cite{Fukushima:1969relu} activation functions. 
Increasing the number of hidden layers also increases the number of parameters contained in the prediction model, but more isn't always better. 
In practice, one has to find a balance between a network that fails to learn due to not enough model parameters and a network that overfits the training data due to too many model parameters.

Using five hidden layers à 512 hidden units, test-set accuracies as high as $56.2 \% \pm 0.3 \%$ could be reached, which is significantly above the $11.1 \%$ accuracy of a random guess. 
Note that we kept the size of our training data set large enough to not easily overfit a model of this size.
Still, however, approximately every second prediction of this classifier on the synthetic data is wrong.
In Section \ref{combination}, a classifier with a much higher accuracy will be presented.

\paragraph{Regressor}
After the first step of identifying the pole configuration of any particular input via the ANN classifier described above, the second step is to determine the coefficients of the poles in the sum.
To achieve this, we trained another ANN in the form of a regressor.
Just like the classifier, the ANN regressor takes the pole curve, sampled at our usual momentum-squared grid points as its direct input. 

It is worth noting here that a combined loss function was used to train the regressor.
In a weighted sum, the first part of the loss function consisted of a simple mean-squared error (MSE) between the exact and the predicted pole parameters, while the second part was formed by a reconstruction loss comparable to the loss functions used for the training of autoencoders \cite{Rumelhart:1986ae,Wang:2014ae}. 

An autoencoder is a form of ANN where the input data is first fed through several layers with reducing size such that a short or compact \emph{encoding} results. 
From this encoding, a second part of the network with layers increasing in size produces data of the same size and structure as the input, in fact attempting to reconstruct the input as closely as possible.
This is achieved by defining a loss function that punishes deviations of the output from the input such as an MSE or another suitable metric.

In the reconstruction, the predicted pole parameters from the ANN's output were used to calculate the corresponding pole sum's data curve. 
The MSE between this calculated curve and the original (input) curve formed the second part of the loss function.
Architecture-wise, a feedforward neural network with six hidden layers à 256 hidden units was used.

\paragraph{Comparison of performance}
In order to compare the performance of the ANN regressor to the conventional methods, it is easiest to focus on a single pole class.
Herein, $N_r=0$ and $N_c=3$ is used as such an example.
Analogous regressor training and the corresponding analysis are possible for all other classes in the same fashion as outlined in the example.

\begin{figure}[ht]
\centering
\includegraphics[trim=50 40 90 60bp,clip,width=1.0\linewidth,keepaspectratio]{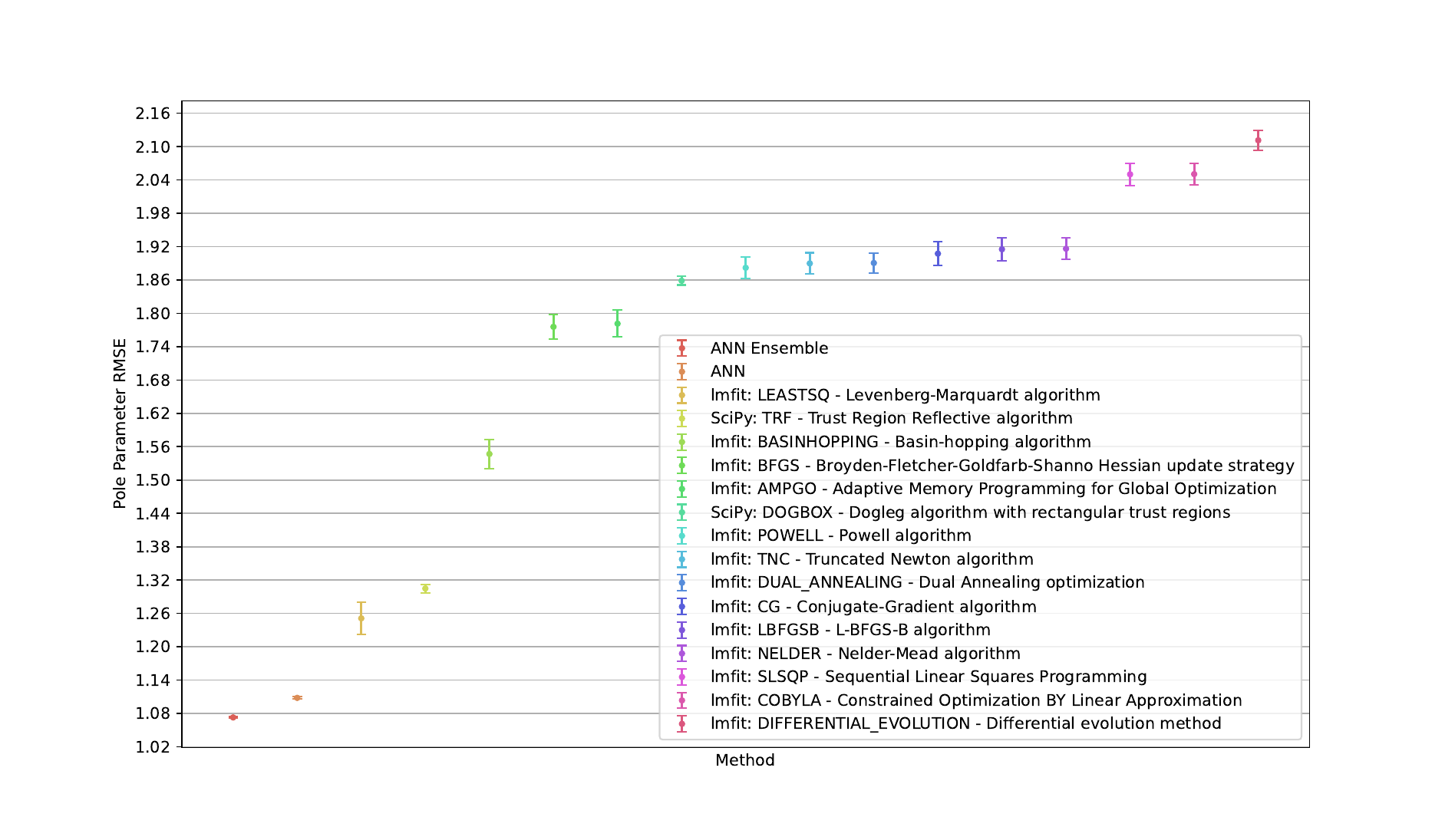}
\caption{
Comparison of results for different fitting algorithms including bounded conventional and ANN methods. See text for discussion.}
\label{fig:comparison}
\end{figure}

Thus, for the comparison of all methods, synthetic data curves generated from pole sums of three complex-conjugated pole pairs were used as inputs to the various regression/fitting methods.
In each case, performance is measured via finding the pole parameters corresponding to each curve most exactly.
In Tab. \ref{tabcompareregression} and Fig. \ref{fig:comparison} the root-mean-squared errors (RMSEs) of the pole parameters predicted by the various methods are presented as they were obtained on synthetic data for the groups of algorithms from the SciPy and LMFIT Python libraries, respectively, as listed in Sec.~\ref{standard}. 

\begin{table}[!ht]
\caption{Performance of an ANN regressor and three conventional fitting methods on synthetic data. The RMSE values measure the difference between the exact and the predicted/fitted pole parameters.}	\label{tabcompareregression}
	\begin{center}
		\begin{tabularx}{0.7\columnwidth}{lc}\hline
			Method  & RMSE  \\\hline 
			ANN  & $1.108 \pm 0.002$  \\ 
			ANN Ensemble  & $1.073 \pm 0.001$  \\
			Unbounded LM  & $0.054 \pm 0.008$  \\
			Bounded TRF  & $1.305 \pm 0.008$  \\ 
			Bounded DOGBOX  & $1.859 \pm 0.008$  \\ 
			\hline
		\end{tabularx}
	\end{center}
\end{table}

As can be seen from Fig. \ref{fig:comparison}, the ANN regressor outperforms all tested bounded methods.
Moreover, training five ANN regressors (same architecture) and averaging their predictions into an ANN ensemble \cite{Perrone:1985en} model improves performance even a little further. 

On the other hand, unbounded LM massively outperforms the ANN (see Tab.~\ref{tabcompareregression}).
However, the downside of unbounded LM when used for this particular problem is its lower probability of converging inside the parameters boundaries compared to the other methods.
More precisely, on synthetic data in $0.58 \% \pm 0.02 \%$ of all fitting attempts with this algorithm, the fitted parameters converged outside the parameter boundaries specified in this section. 
In these ``failed fits'', unbounded LM tends to return very large values for the pole positions and pole coefficients, which lie far outside the domain of physically relevant boundaries. 
For a reminder of these boundaries, please refer to Fig.~\ref{fig:scatterplot} in Sec.~\ref{standard}.

\section{Combining standard approaches and ANNs}\label{combination}

\paragraph{CFNN classifier}
Since this paper's investigation is not just about comparing different methods, but also about finding the best solution to our original use case, we proceed to investigate the combination of ANN methods with conventional fitting techniques.
As mentioned above in Sec.~\ref{anns}, this approach helps us to arrive at a classifier with a much higher accuracy than the ANN-only version described there.

Basically, there are two variants of such a combination approach. 
First of all, one can use ANN predictions as initial conditions in conventional fitting techniques. 
Secondly, one can use fit results as (part of the) ANN inputs.
Our curve-fitting plus neural network (CFNN) classifier follows the second route and combines information gathered from conventional fitting methods with an ANN in order to predict the pole class of the data curve, that is, $N_r$ and $N_c$.
In more detail, the CFNN classifier is constructed in the following manner:

\begin{enumerate}
\item In a first step, a set of conventional fitting methods is chosen. 
A single method can also be chosen and used multiple times.
\item For each entry in the list created above and for each combination of $N_r$ and $N_c$ the pole sum is fitted to the data. 
Each fit returns a list of fitted parameters.
Additionally, for each fit, an MSE between the original data curve and a data curve created from the fitted parameters can be calculated.
\item All information gathered in the previous step, that is, all the fitted parameters and calculated MSEs are then concatenated into a single large vector. 
This vector is then used as the input vector for an ANN classifier. 
This is in contrast to Sec.~\ref{anns}, where the data curves themselves were used as inputs for the ANN.
\end{enumerate}

Using a simple feedforward architecture with two hidden layers à 64 hidden units and the set $\{$ unbounded LM, unbounded LM, bounded TRF, bounded DOGBOX, bounded DOGBOX $\}$ of fitting methods, a test accuracy of $97.31 \% \pm 0.01 \%$ on synthetic data was achieved.
In other words, the CFNN classifier massively outperforms the direct classifier presented above in Sec.~\ref{anns}.

\paragraph{Binary CFNN classifier}
In the previous paragraph, we presented a direct approach to make use of both ANN and conventional fitting methods in order to solve the general classification problem for our use case.
However, this is not the only question one can ask to get information about the analytic structure of the quark dressing (or other) functions based on real-axis-only information.

In cases where the maximum number of poles (i.e., the upper limit for $N_r + N_c$) can not be estimated or is supposedly too high for feasible calculations (or for some other reason), one may resort to a simpler strategy.
In practice, it might be sufficient to analyze the pole configuration closest to the origin (in terms of the position's absolute value).
In the following, we refer to this pole (for real position) or pole pair (for complex conjugate positions) by simply writing the \emph{first pole} for the sake of brevity.
 
In order to predict whether the first pole's position is real or complex (with a conjugated partner), a binary classifier can be trained.
We constructed such a binary classifier in the same way as the CFNN classifier presented above.
The synthetic data for training and testing purposes was created using the following boundaries:

\begin{itemize}
\item $N_r + N_c \leq 3$
\item $ \Re\left(v_n\right) \in \left[-2,-0.05\right], \Im\left(v_n\right) \in \left[0,2\right]$
\item $w_n \in \left[-2,-0.05\right]$
\item $\Re\left(c_n\right) \in \left[-1,1\right], \Im\left(c_n\right) \in \left[-1,1\right]$
\item $d_n \in \left[-1,1\right]$
\end{itemize}

Using a feedforward ANN with two hidden layers à 16 units and $\{$ unbounded LM $\}$ as the set of fitting methods, a test accuracy of $98.44 \% \pm 0.04 \%$ on synthetic data was achieved.
Since the unbounded LM algorithm showed convergence problems with real-world data, another version of this binary classifier, using $\{$ bounded TRF, bounded TRF, bounded TRF $\}$ as the set of fitting methods, was trained and is used in Section \ref{application}.
For comparison: using a single hidden layer and 64 hidden units resulted in a test accuracy of $94.80 \% \pm 0.06 \%$.

\paragraph{Combined Classifier-Regressor Approach}
Finally, we present a combined approach to solve the complete problem, that is, first determine the number of poles and then determine their pole parameters. 
Combining the results from this section and Sec.~\ref{anns}, the following approach can be seen to result in the highest accuracies and chance of converging within the selected parameter boundaries (on synthetic data):

\begin{enumerate}
\item Use a CFNN classifier to determine $N_r$ and $N_c$.
\item Use the predicted values of $N_r$ and $N_c$, fit the pole sum to the data using unbounded LM. 
\item Repeat this step multiple times to create a scatter plot, like Fig.~\ref{fig:scatterplot}, in order to make sure that the predicted pole parameters are not heavily dependent on the initial guesses.
\item If unbounded LM converges within the desired boundaries and the scatter plot does show only little dependence of the fitted parameters on the initial guesses, then the fit can be considered successful.
\item Otherwise, use an ANN regressor to predict the pole parameters.
\end{enumerate}

In an obvious extension to this, one might try to find new or better suited conventional fitting methods.
These could then not only be used in creating the CFNN classifier's input vector and lead to even higher classification accuracies, but may also replace unbounded LM (or the ANN regressor) in the approach presented above.

\section{Open-source implementation}\label{code}
The source code and corresponding instructions can be found in the project's GitHub repository at:
\url{https://github.com/siegfriedkaidisch/Pole_Sum_Net}

\section{Real-world application}\label{application}

\paragraph{General setup}
Since this methodical improvement was motivated by a concrete use case in theoretical hadron physics, we want to also test it for this particular setup. 
In order to do this, we produced several sets of real-axis data for the quark dressing functions for a number of commonly used model interactions in the quark DSE.
To obtain these, we solved the quark DSE numerically via standard iterative techniques, and input the solutions into the various classifiers and the regressors described above. 

It is interesting to note here that the dominant singularity structure is known from direct numerical continuations into the complex plane for some of the standard interactions. 
As already mentioned in the Introduction, this can be achieved for special cases via a suitable numerical implementation of complex path integration.
In particular, there are two different cases (one complex-conjugate pole pair and two real poles) that we can readily test with rather standard values of the corresponding model parameters. 
In addition, a prediction can be made for a modification of the dressed-quark-gluon-vertex structure that is not accessible to the same methods easily.
Last but not least, we also discuss our predictions with respect to the particular model choices in Ref.~\cite{Bhagwat:2002tx} with the three-complex-pole-pair fits made therein.

In order not to clutter the main text of this paper with too many technical details, these have been deferred to the appendix. 
In \ref{technicalities}, we first present the rainbow approximation of the quark DSE in detail, and derive a functional form that is defined by a standard set of parameters.
Next, we collect a list of model parameters and assign a name to each parameter set for easy reference in the main text.
For the purposes of this section, it is not necessary to refer to each specific set of parameters, but naturally, the interested reader can make this reference at any time.
A summary of the parameter sets and models used for testing can be found in Tab.~\ref{tablemodelparams}.

\paragraph{Classification of location and number of singularities}
The first concrete problem to focus on is the application of our classifier for finding the pole configuration. 
In order to arrive at a concrete prediction for each particular case/parameter set and get an error estimate at the same time, we trained an ensemble of 100 classifiers and combined predictions from each into a histogram of predicted classes.
Here, a reminder about the ANNs used in this section is in order: 
in Sections \ref{anns} and \ref{combination}, the ANNs were trained on and applied to synthetic data. 
In contrast, all ANNs used in this section were trained on synthetic data, but are applied to real-world data. 
Given the differences between synthetic and real-world data, some of which are outlined here (maximal number of poles, numerical artifacts, more complex pole types, etc.), the ANNs will necessarily show worse performance in this section than demonstrated above.

The class numbers and the corresponding configuration are given in Tab.~\ref{tab:classes}.
\begin{table}[htp]
\caption{Numbers of real and complex-congugate poles corresponding to each class.}
\begin{center}
\begin{tabular}{ccc}\hline
Pole class & real poles & complex conjugate pairs \\\hline
0 & 1 & 0 \\
1 & 0 & 1 \\
2 & 2 & 0 \\
3 & 1 & 1\\
4 & 0 & 2 \\
5 & 3 & 0 \\
6 & 2 & 1 \\
7 & 1 & 2 \\
8 & 0 & 3 \\\hline
\end{tabular}
\end{center}
\label{tab:classes}
\end{table}

\begin{figure*}
	\centering
    \subfloat[M-BV-1]{%
        \includegraphics[width=0.19\textwidth]{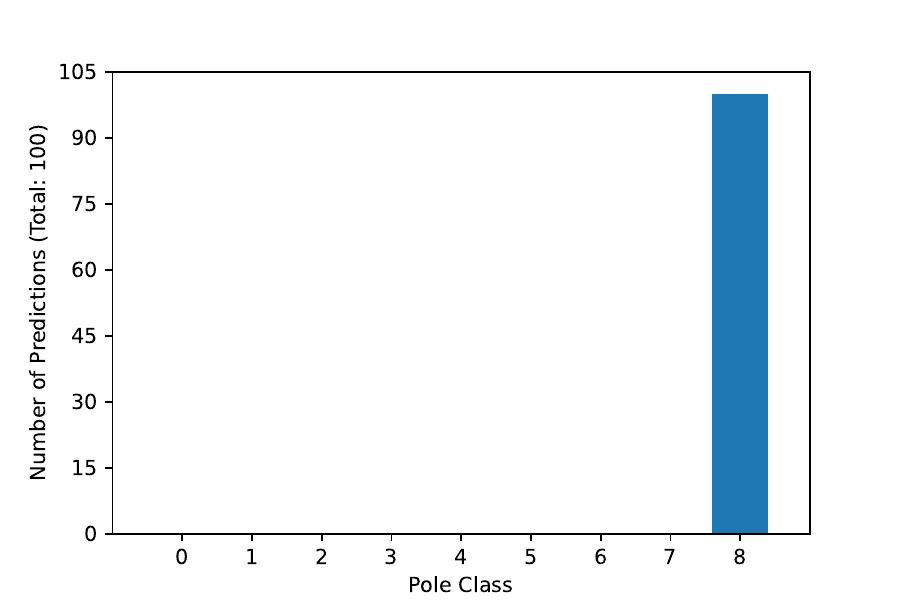}
        \label{fig:sub1}
    }
    \hfill
    \subfloat[M-BV-2]{%
        \includegraphics[width=0.19\textwidth]{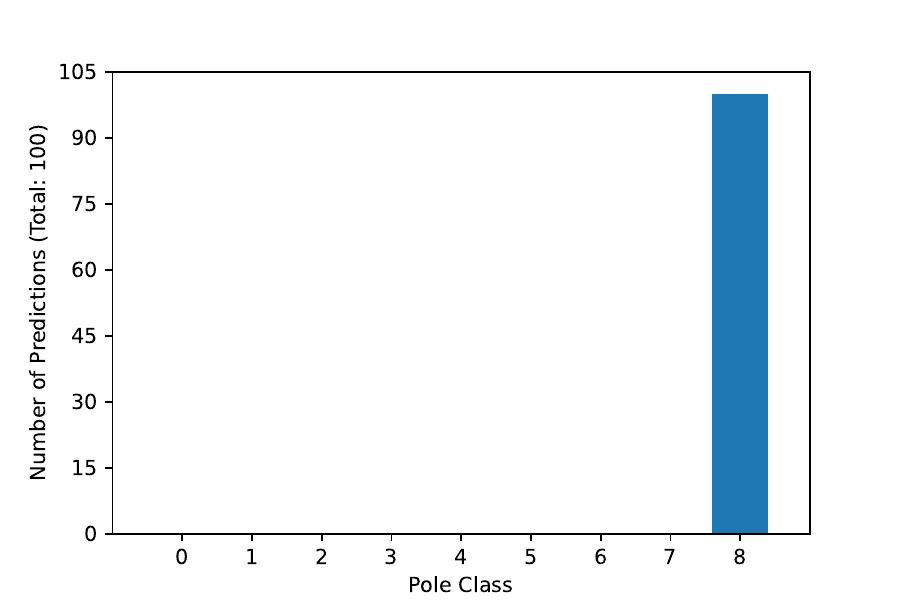}
        \label{fig:sub2}
    }
    \hfill
    \subfloat[M-BV-3]{%
        \includegraphics[width=0.19\textwidth]{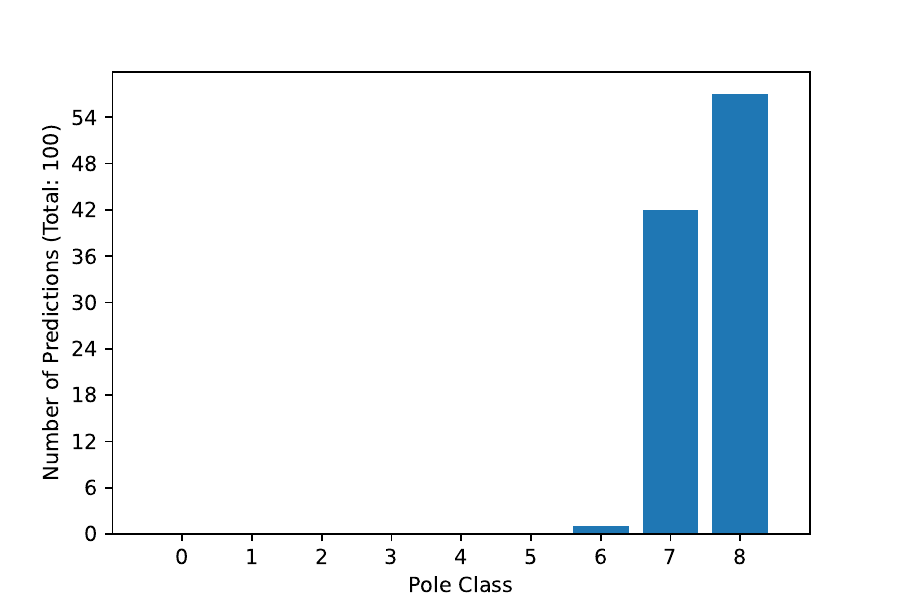}
        \label{fig:sub3}
    }\hfill
    \subfloat[A-BV-1]{%
        \includegraphics[width=0.19\textwidth]{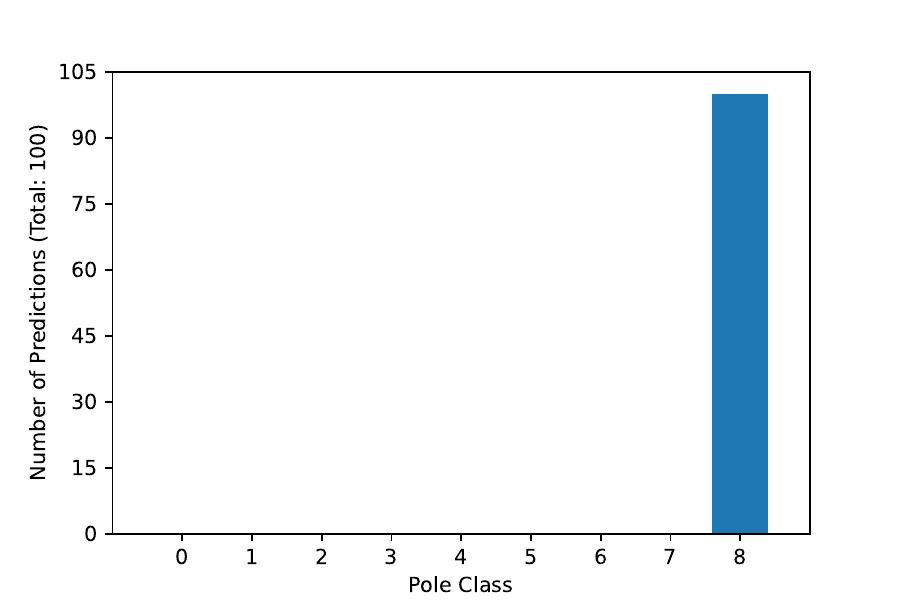}
        \label{fig:sub4}
    }
    \hfill    
    \subfloat[A-BV-2]{%
        \includegraphics[width=0.19\textwidth]{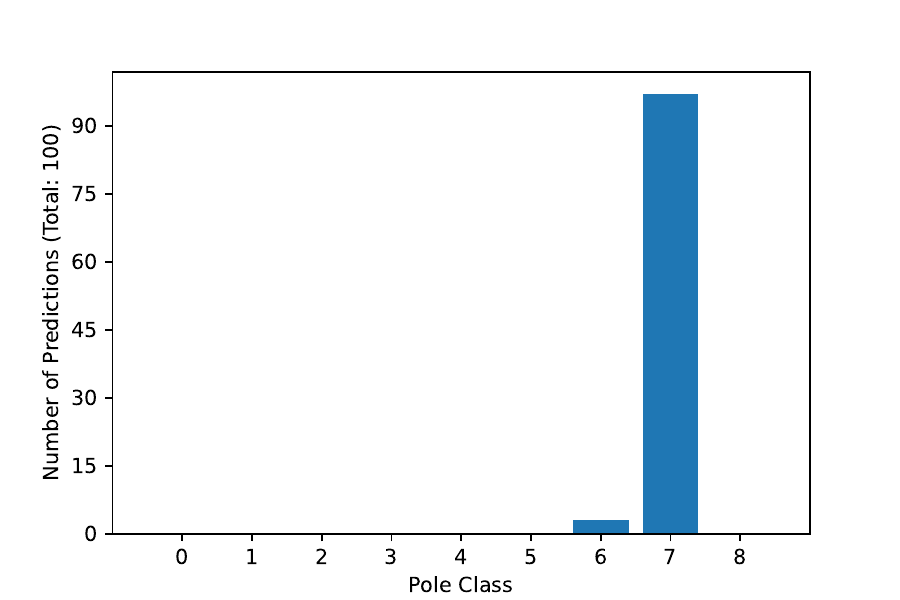}
        \label{fig:sub5}
    }\\ 
    \subfloat[M-BV-3 (closeup)]{%
        \includegraphics[trim=170 0 90 20bp,clip,width=0.23\textwidth]{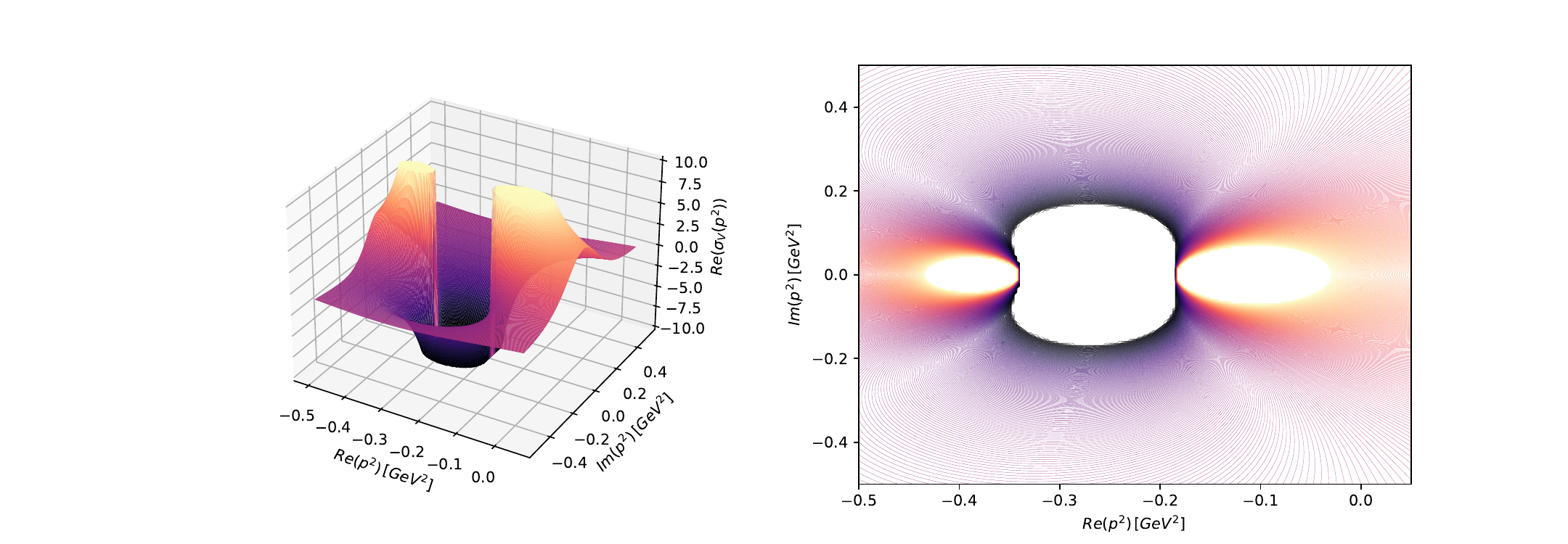}
        \label{fig:sub6}
    } 
    \hfill
    \subfloat[M-BV-3]{%
        \includegraphics[trim=170 0 90 20bp,clip,width=0.23\textwidth]{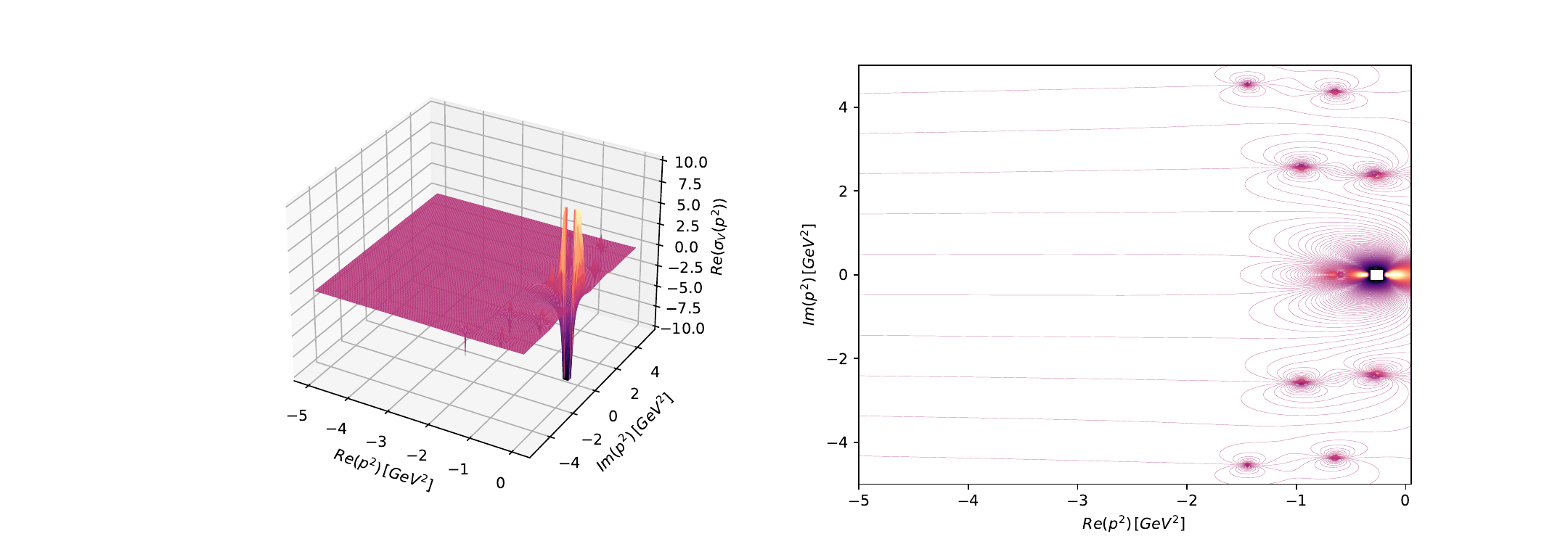}
        \label{fig:sub9}
    }	
    \hfill  	
    \subfloat[A-BV-1 (closeup)]{%
        \includegraphics[trim=170 0 90 20bp,clip,width=0.23\textwidth]{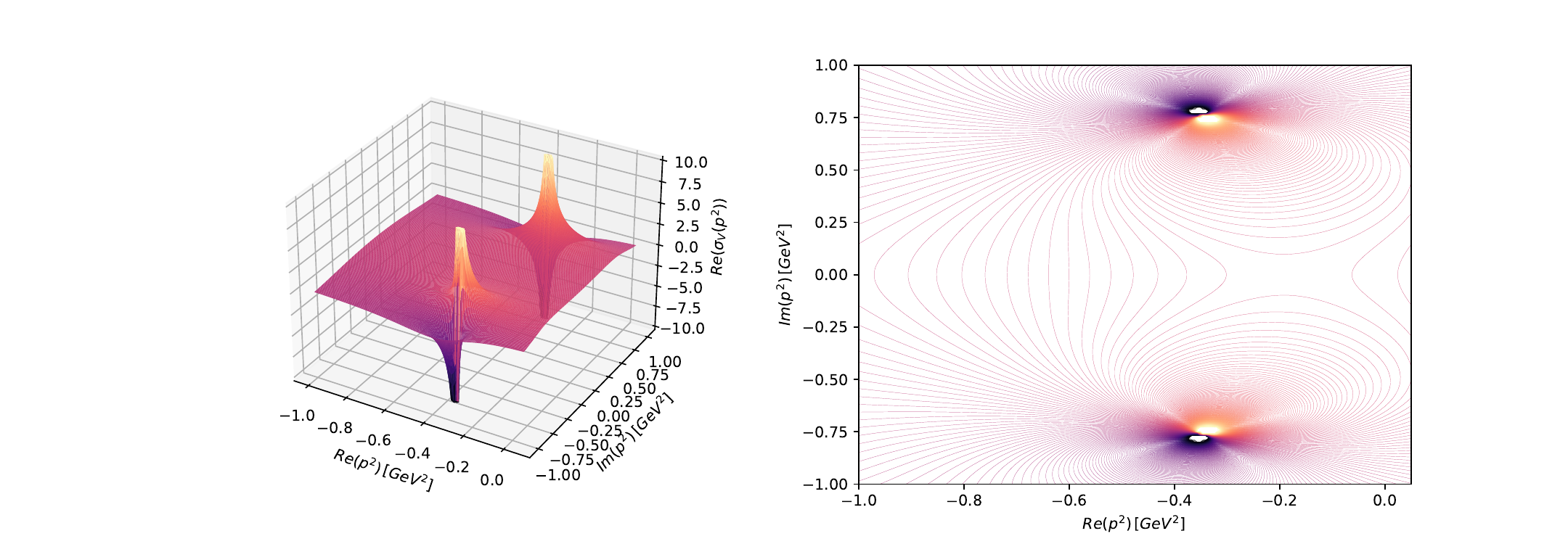}
        \label{fig:sub7}
    }
    \hfill    
    \subfloat[A-BV-2 (closeup)]{%
        \includegraphics[trim=170 0 90 20bp,clip,width=0.23\textwidth]{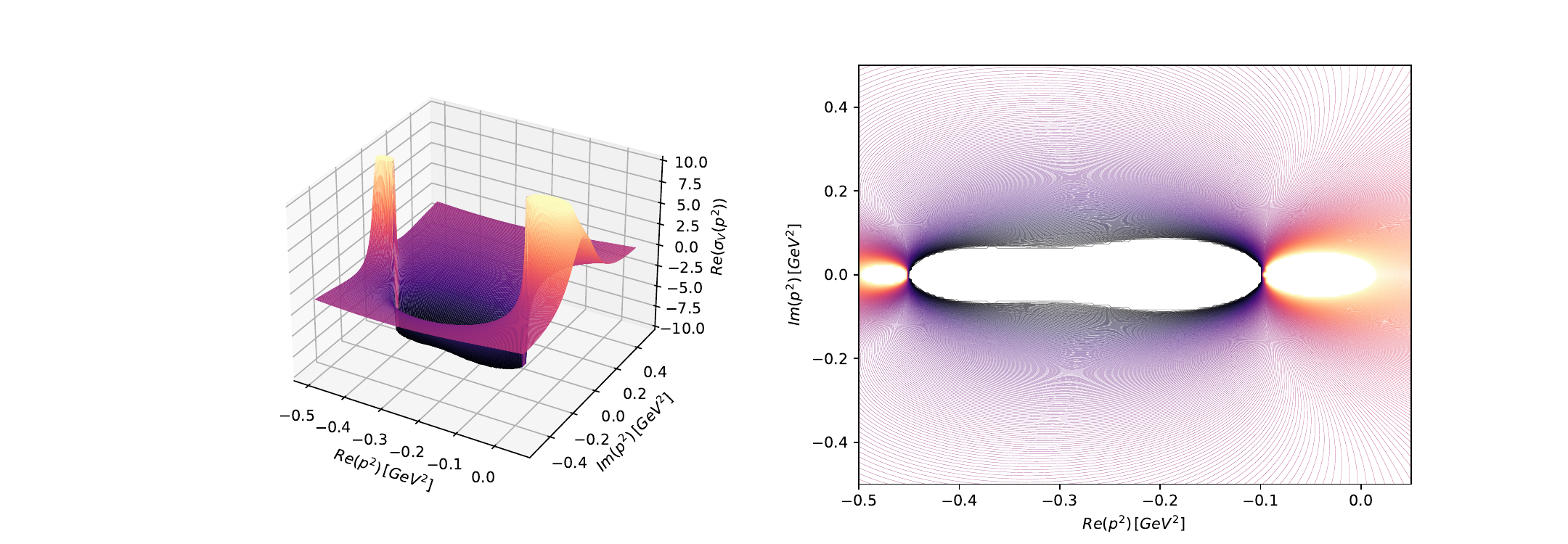}
        \label{fig:sub8}
    }
	  \caption{\emph{First row}: Histograms for predicted classes for 5 different interaction models. \emph{Second row}: complex-plane renderings of real parts of dressing functions for different model interactions. For M-BV-3 and A-BV-2, the dominant poles are real. For all others, the dominant poles are complex conjugate.
	  \label{fig_numeric_continuation}}
\end{figure*}

The results for the histograms for our five bare-vertex model setups are shown in Fig.~\ref{fig_numeric_continuation}, subplots (a)--(e).
The predictions by majority are mostly clear except for M-BV-3 shown in subplot (c), where classes $7$ and $8$ are predicted with similar probability.

The interesting point in this situation is that we know from illustrative numerical continuation into the complex plane, which configuration is actually present.
To this end, we focus on the region close to the origin and ignore numerical artifacts to the best of our ability.
In essence, there are two possible scenarios: one, where the dominant pole is real, and another, where the dominant structure is a complex-conjugate pole pair.
Subplots (f)--(i) illustrate the actual situation in the complex plane corresponding to some of the model cases.

Apart from the dominant poles (closest to the origin), we observe a number of weaker poles farther away from the origin for all cases of model studies, as exemplified for M-BV-3 in subplot (g).
What this implies is that, as noted in the beginning of the description of our approach, our 9 classes are only an approximation to more complicated situations present in real-world data.

Still, keeping these circumstances in mind, we can interpret our predictions and the comparison to the actual situations as follows:
\begin{itemize}
\item In all tested cases, the classifiers always predict classes $6$, $7$ or $8$. 
This shows that the classifiers detect the additional structure in the dressing functions, i.e. the additional poles farther from the origin, which is to be expected by virtue of how model training was actually conducted.
\item In scenarios where there are no dominant real poles, the prediction is class $8$, which corresponds to three (the maximum in our case) pairs of complex-conjugate pole pairs.
\item In scenarios where there are dominant real poles, like in M-BV-3 and A-BV-2, we observe mixed results: the data from the numerically less troublesome interaction (A) produces the clearer prediction (e), while the result for (M) is mixed. Actually, it is mostly at odds with the actual situation in that it fails to account for the dominant real-pole structures and predicts these as part of the scenario in less than half of the ensemble predictors.
\item Being aware of these limitations, we could try to take predictions of any class other than $8$ as a signal for some other structure than a sum of complex-conjugate pole pairs. However, we do not find these results very trustworthy, because they are not reliable, either. Consider subplot (e), where, in terms of dominant structures, class $6$ would have been a better outcome. The class-$7$ prediction, however, points to a single real pole instead.
\item Overall, we conclude that on real-world data, our predictors could serve as reasonable indicators, if one would enlarge the number of classes and/or include the pole strengths (their residues) as training weights. We suggest to include these points in future studies.
\end{itemize}

\paragraph{Regression for pole positions}
The second concrete problem is the application of our regressor for finding pole positions, once their configuration has been predicted.
More precisely, the regressor estimates the parameters for the predicted class in terms of the pole configuration.

While one could make the assumption that the predicted class is correct, and then estimate the pole positions based on the corresponding regressor prediction alone, it is also instructive to actually consider the regressor predictions for all classes and compare them to those cases, where the positions of the dominant poles are known, as shown above.
It is important to note here that both regressor and traditional fitting methods should be used to find a good estimate for real-world data.
In the following, we analyze an illustrative sample from our results in order to show the facets and the possibilities of predictions with both kinds of tools.

Recalling the idea that the dominant singularity would be most important in a numerical application of our regressor, we start with comparing the prediction values for different pole configurations for a single model and parameter set.
This makes sense because the lower classes correspond to emphasizing the dominant poles only and neglecting the rest, while the higher classes take more poles into account.
Thus, in our results we can see the difference between a fit with a few parameters only that tries to capture the dominant contribution in the function in question and a fit that is allowed additional parameters to try and capture smaller corrections in such a way that the dominant contribution can be fit more accurately.

However, our results show that this is not always what happens.
In Tab.~\ref{results:regressor} we have collected the real and imaginary parts of the dominant pole positions for the M--BV--2 model, together with the actual positions extracted from numerical continuations, as introduced above.

\begin{table}[!ht]
	\caption{Predictions for pole positions for several different predictors, based on four different classes and both regressor- and TRF-predictions, versus the actual values (first line) for model M--BV--2}	\label{results:regressor}
	\begin{center}
		\begin{tabularx}{\columnwidth}{lccrrc}\hline
			Setup & Class & $\Re_1$  & $\Im_1$  & $\Re_2$ & $\Im_2$  \\ \hline
			Cont. & -- & $-0.211$ & $0.343$ & $-1.09$ & $0.48$ \\
			TRF-fit & 1 &  $-0.271$ & $0.336$ & -- & -- \\
			Regressor & 1 &  $-0.212$ & $0.184$ & -- & -- \\
			TRF-fit & 4 & $-0.217$ & $0.344$ & $-0.868$ & $1.198$ \\
			Regressor & 4 & $-0.209$ & $0.275$ & $-1.420$ & $0.756$ \\
			TRF-fit & 7 & $-0.188$ & $0.357$ & $-0.105$ & $1.207$ \\
			Regressor & 7 & $-0.288$ & $0.323$ &  $-1.363$ & $0.892$ \\
			TRF-fit & 8 &  $-0.204$ & $0.358$ &  $-0.747$ & $1.517$ \\
			Regressor & 8 &  $-0.220$ & $0.346$ & $-0.692$ & $0.492$ \\\hline
		\end{tabularx}
	\end{center}
\end{table}

In these results, we see that the predictions are of mixed quality.
While all of them are in the general ballpark, and the results tend to get better the higher the class, some differences can be seen for the real and imaginary parts of each pole as well as for first and second pole.
Furthermore, the TRF-fit and the regressor differ in how accurately the value found in the continuation is reproduced.
In summary, no particular pattern emerges as to which method could be superior based on these results alone.

\begin{figure}[ht!]
\centering
    \subfloat[Predicted results]{%
	\includegraphics[width=1.0\columnwidth,keepaspectratio]{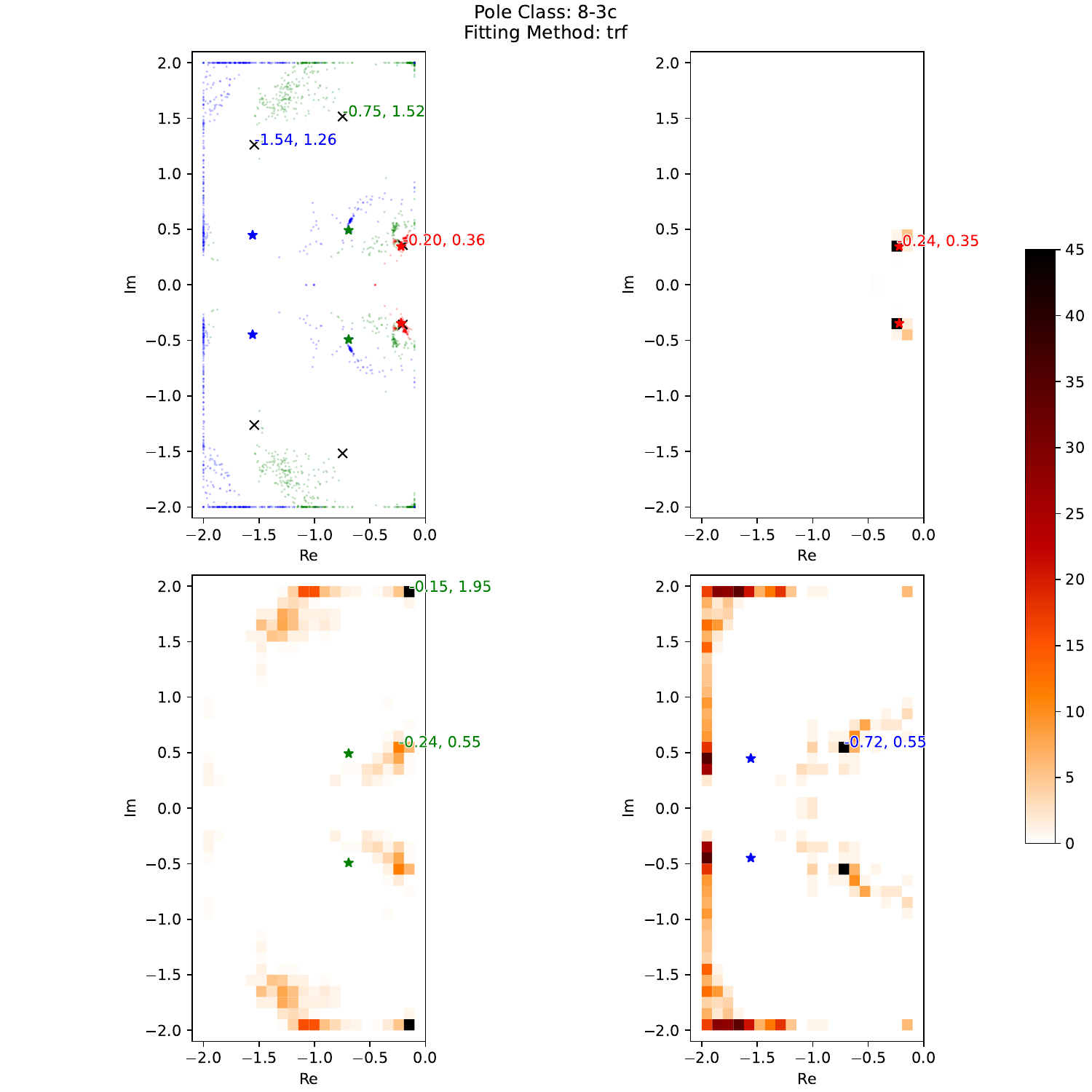}
        \label{fig:mbv2num}
    }\\
    \subfloat[Numerical continuation (closeup)]{%
        \includegraphics[width=\columnwidth]{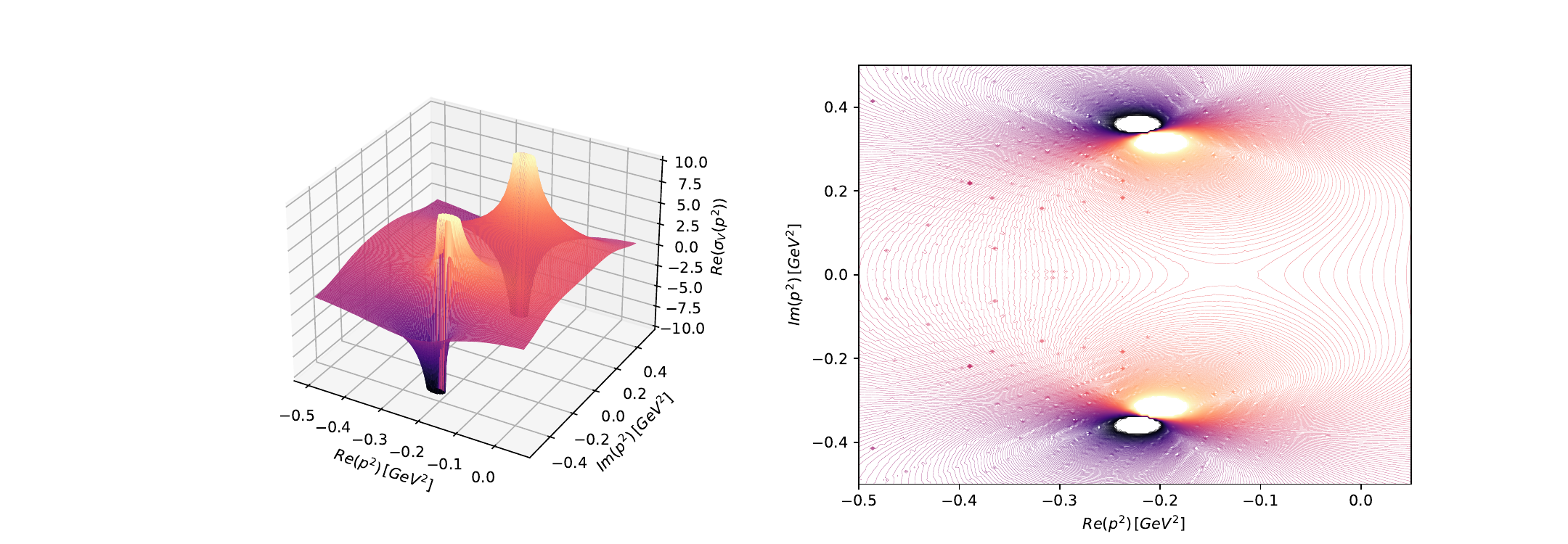}
        \label{fig:mbv2close}
    }\\
    \subfloat[Numerical continuation]{%
        \includegraphics[width=\columnwidth]{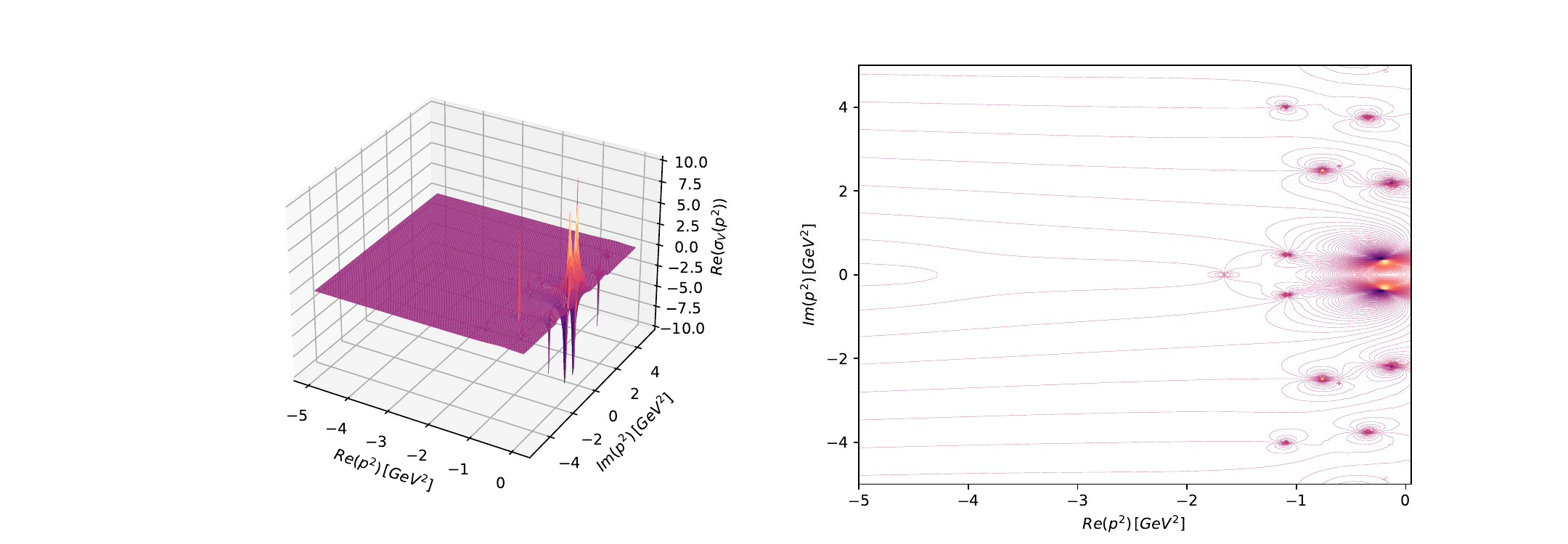}
        \label{fig:mbv2}
    }
\caption{
(a) Visualization of the predictions for pole positions for class $8$ obtained for model M--BV--2. For a detailed description, see the text. (b) and (c) Closeup and wide visualization of numerical continuation for this model. }
\label{fig:classpredictions}
\end{figure}

In order to understand better where these numbers come from, we also provide illustrations for all of the classes for this model.
As an example, the predictions are presented in Fig.~\ref{fig:classpredictions} (a) for class $8$ and model M--BV--2.
The same figures for all classes and this model are presented in \ref{app:classpredictions} for a comprehensive presentation.
In such a figure, two kinds of subplots are combined: 
\begin{itemize}
\item The first subplot shows the combination of all predictions into one figure, together with visualizations of the resulting regressor- and TRF-predictions.
\item The second kind of subplot shows a visualization of the analysis for each pole (pair) separately.
\item As a result, a different number of subplots of the second kind accompany the first kind in each case, according to the number of different poles in each class.
\item In Fig.~\ref{fig:classpredictions} (a) for class $8$, which corresponds to three complex-conjugate pole pairs, we thus have three subplots of the second type, corresponding to the first, second, and third pair of poles.
\end{itemize}

In the first subplot, the point clouds represent individual TRF-fit results based on different initial conditions in the fit.
The colors correspond to the different singularities within one prediction, i.\,e., for class $8$ the positions of the first, second, and third pair of complex-conjugate poles. 
The stars in the figures correspond to the corresponding predictions of our regressor.
The black symbols \textbf{X} mark the averages of the TRF predictions, annotated with the concrete numbers for the real and imaginary parts.
Those averages are the values found in Tab.~\ref{results:regressor} in the "TRF-fit" rows.

In the second kind of subplot, where each analysis is visualized separately, we show 2D-histograms of the individual TRF predictions via the color code to the right of the figure.
The stars for the regressor predictions are present here as well.
In addition, the color-coded areas in these plots indicate the weight of the distribution in different areas in the complex plane.
The numbers indicate the position of the square with the highest weight, which can be interpreted as the mode of the distribution.
Note that some accumulation happens at the boundary of the plot: this is due to the restrictions on the TRF fit, which converges towards or at that boundary for some of the initial conditions in the fit.
For this reason, if the mode is found to lie on the boundary, additionally the mode of all fits that did \emph{not} converge at or near the boundary is given in the subplot numerically, as is the case in the subplot in the bottom left of Fig.~\ref{fig:classpredictions} (a).
	
As a final comment on this problem, we would like to stress the potential of the combined approach again, where information from both the TRF fits and the regressor model are combined by using suitable tactics. 
In order to see the comparison to the actual numerical continuations into the complex plane, we have added visualizations thereof in subfigures \ref{fig:classpredictions} (b) and \ref{fig:classpredictions} (c).

\paragraph{Classification of dominant-singularity location}
In the previous two subsections, we have shown how our results may guide an investigation of data indicating a complicated analytical structure in the complex plane.
However, we have seen that a direct interpretation is not always easy, in particular, if concrete and reliable results are to be obtained.

In a more practical approach, we want to highlight the possibility to predict the configuration of the dominant poles in a given case. 
For our intents and purposes, the dominant pole is located closest to the origin and in the domain of negative real parts of the momentum squared.
The reason for investigating this is that, for many applications, it is a good first approximation to take into account only the dominant pole contribution and neglect the rest.

To this end, we use the binary CFNN classifier described above, which essentially predicts one of two classes, namely:
\begin{itemize}
\item The dominant singularity is a pole on the negative real axis (referred to as class $0$ in this case).
\item The dominant singularity is a pair of complex-conjugate poles (referred to as class $1$ in this case).
\end{itemize}
To obtain the predictions, we use an ensemble of 100 classifiers, as before, and report the overall prediction in the form of the percentages of class-$0$ and class-$1$ predictions in the ensemble.
The results of this analysis are summarized in Tab.~\ref{tab:binary} and contrasted to the actual situation known from the numerical continuation.
We make the comparison for all models with the bare vertex in the quark DSE.

\begin{table}[!ht]
	\caption{Predictions for dominant pole contributions based on ensemble predictions from our binary CFNN classifier, compared to the actual class obtained via the continuation.}	\label{tab:binary}
	\begin{center}
		\begin{tabular}{lcrr}\hline
			Model & Cont. & Prob.~$0$  & Prob.~$1$   \\ \hline
			M--BV--1 & $1$ & $0.00$ & $1.00$ \\
			M--BV--2 & $1$ & $0.00$ & $1.00$ \\
			M--BV--3 & $0$ & $0.16$ & $0.84$ \\
			A--BV--1 & $1$ & $0.00$ & $1.00$ \\
			A--BV--2 & $0$ & $0.95$ & $0.05$ \\\hline
		\end{tabular}
	\end{center}
\end{table}

From this table, we see that the ensemble is pretty confident in its predictions in the sense that probabilities are either $1$ and $0$, or clearly indicate one class over the other.
In comparison to the actual configuration as indicated in the \emph{Cont.} column, we can see that the only wrong prediction is a class $1$ instead of the actual $0$, which also has the lowest confidence over the models investigated here.

Another observation is that all actual classes $1$ (complex-conjugate pole pair is dominant) are predicted correctly and with full confidence.
For the actual classes $0$ (real pole is dominant), the prediction is correct for the model with the smaller amount of numerical artifacts (A--BV--2), while the prediction is wrong for the numerically more troublesome model M--BV--3 (also with somewhat lowered, but still high confidence).

In preparation for the following analysis of the Ball-Chiu-vertex versions of these models, we summarize these findings in this rather specialized situation as follows:
\begin{itemize}
\item If we find a $100$-percent prediction for class $1$, it tends to be correct.
\item If we find a prediction with high confidence for class $0$, it is probably correct.
\item If we find a prediction with lower confidence for class $1$, it is probably wrong.
\item The latter in particular applies to situations that are numerically involved (M models rather than A models).
\end{itemize}

We would like to note that our analysis has not been exhaustive by means of varying models, parameters, and changing other circumstances of the setup, but serves as a first guide to the potential of the method.
More detailed investigations are beyond the scope of the present paper and should be performed in future studies.

\paragraph{Predictions for the Ball-Chiu vertex}
After this analysis of known real-world data and model situations, we want to add our take on predicting the analytic structure caused by a non-trivial version of the dressed quark-gluon vertex in Eq.~(\ref{eq:dse}).
The point of this exercise is that vertices beyond the bare vertex are necessary in realistic models of hadrons in QCD.
Furthermore, such an analysis can be performed for dressing functions of other more complicated Green and vertex functions as well, where a numerical continuation into the complex plane may be prohibitively complicated.

In this sense, we implemented a rather immediate application of our method to something new.
Concretely, we investigate a well-known model for the dressed quark-gluon vertex in Eq.~(\ref{eq:dse}) proposed some time ago by Ball and Chiu \cite{Ball:1980ay}. 
The essential information for the purpose of the present work is collected in \ref{bcv}, in particular the functional form of the Ball-Chiu vertex, which is given in Eq.~(\ref{eq:bcv}) for our notation.
While this particular vertex model is rather easy to use on the positive real axis of the momentum-squared plane, it is not straight-forward to use numerical methods for a continuation into the complex plane.

That said, in applications where the quark dressing functions are required for complex arguments, an ansatz using information from results from our method can be sufficient as a first approximation. 
In particular, using the dominant singularity term only should be reasonable as well. 
In this spirit, we apply the binary CFNN classifier together with the insights gained in the previous subsection in order to try and make sense of what we would expect the dominant singularity in the quark dressing functions to be for a model using the Ball-Chiu vertex construction.

We used the same setup as above to predict the dominant singularity closest to the origin with our binary-CFNN classifier ensemble, which we benchmarked on known-class data (see results in Tab.~\ref{tab:binary}).
In terms of the models used, we kept the basic model setup the same for the five different models. 
However, due to interaction strength added by the nontrivial compared to the bare vertex, we reduced the parameter $D$ in such a way that dressing functions remained reasonable on the positive real axis.
For a list of the corresponding model parameters, please refer to Tab.~\ref{tablemodelparams} in \ref{bcv}.

The prediction results obtained for the Ball-Chiu vertex are presented in Tab.~\ref{tab:binarybc}.
This table is analogous to Tab.~\ref{tab:binary} in the sense that the models correspond to each other, but have the vertex replaced and one parameter adjusted.
Furthermore, the column with the correct class inferred from numerical continuation above has been replaced by a column with the prediction result in terms of the class with higher probability.

\begin{table}[!ht]
	\caption{Predictions for dominant pole contribuations based on ensemble predictions from our binary CFNN classifier for the Ball-Chiu versions of the five models analyzed in the previous subsection.}	\label{tab:binarybc}
	\begin{center}
		\begin{tabular}{lcrr}\hline
			Model & Prediction & Prob.~$0$  & Prob.~$1$   \\ \hline
			M--BC--1 & $1$ & $0.40$ & $0.60$ \\
			M--BC--2 & $1$ & $0.15$ & $0.85$ \\
			M--BC--3 & $1$ & $0.13$ & $0.87$ \\
			A--BC--1 & $1$ & $0.36$ & $0.64$ \\
			A--BC--2 & $1$ & $0.49$ & $0.51$ \\\hline
		\end{tabular}
	\end{center}
\end{table}

In terms of interpreting these results, our first observation is the rather low confidence on all of the predictions.
As a consequence, we can say that these predictions have to be taken with a grain of salt.
If we do want to draw some kind of conclusion, we may try to follow the statements given at the end of the previous subsection:
\begin{itemize}
\item Since none of the predicted confidences is $100$ percent, we cannot use the first argument for a class $1$ situation.
\item Neither, there is a convincing prediction of class $0$ in any of these cases.
\item However, since class $1$ is predicted for all cases and the situation is numerically challenging, i.\,e., numerical artifacts are likely, we may argue by analogy (patterns in the models and the situation) that the likely actual class for all of these cases is $0$.
\end{itemize}
The corresponding result would be that replacement of the bare vertex by a Ball-Chiu ansatz in the gap equation and reasonable adjusting the model parameters leads to a dominant real pole closest to the origin in the complex-plane representation of the quark dressing functions.

\paragraph{Comparison to fits in previous works}
Since part of our starting point in this investigation was a successful use case that was based on a fit using three complex-conjugate pole pairs, in other words, corresponding to our class $8$, we now return to this investigation as the last immediate application of our approach and results.

In Ref.~\cite{Bhagwat:2002tx}, the authors used the model M--BV--2, which we also investigated, since it is one of the standard model-parameter sets used in the literature.
In fact, to our knowledge, it is probably the most widely used particular set of parameters in this regard.
Thankfully, the authors reported the pole positions resulting from their fit, which we repeat here and contrast to the outputs from our setup in Tab.~\ref{tab:bptcomp}.

\begin{table}[!ht]
	\caption{Our predictions and analysis compared to the results published in Ref.~\cite{Bhagwat:2002tx} for model M--BV--2. For a detailed description, see text.}	\label{tab:bptcomp}
	\begin{center}
		\begin{tabularx}{\columnwidth}{l|cc|cc|cc}\hline
			  & $\Re_1$  & $\Im_1$ & $\Re_2$ & $\Im_2$ & $\Re_3$ & $\Im_3$  \\ \hline
			Cont. & $-0.21$ & $0.34$ & $-1.09$ & $0.48$ & $-1.66$ & $0$ \\	
\cite{Bhagwat:2002tx} & $-0.21$ & $0.33$ & $-1.28$ & $1.25$ & $-2.13$ & $1.77$ \\	 
			TRF & $-0.20$ & $0.36$ & $-0.75$& $1.52$ & $-1.54$& $1.26$  \\
			Reg. & $-0.22$ & $0.35$ & $-0.69$& $0.49$ & $-1.56$ & $0.45$  \\ \hline 
		\end{tabularx}
	\end{center}
\end{table}

In this table, the columns labeled $\Re_i$ and $\Im_i$ contain values for real and imaginary parts of the position of pole pair number $i$.
The rows contain the following data: 
\begin{itemize}
\item Cont.~refers to the values from the numerical continuation, which correspond to the actual pole positions.
\item \cite{Bhagwat:2002tx} refers to the values from this reference, obtained via a fit to three complex-conjugate pole pairs.
\item TRF refers to our TRF-fit average taken from the same setup.
\item Reg.~refers to our regressor prediction for the class-$8$ configuration.
\end{itemize}

It is noteworthy that the first pole pair is well reproduced in all three approaches.
The second pair less so, and the third pair misrepresents the type of singularity (which in our continuation appears to be on the real axis).
Still, one can also see that in particular the regressor performs surprisingly well for $\Im_2$ and $\Re_3$, and keeps $\Im_2$ rather small.
The main and somewhat obvious observation from comparing parameters beyond those of the dominant pair is that the multi-pole-pair fit is unable to represent a structure that it de facto does not contain (any kind of stronger poles along the negative real axis).

That said, it is interesting to see that the pole positions resulting from fits like these are of similar magnitude but can hint to quite different positions.
This is an important observation concerning studies like the one in \cite{Bhagwat:2002tx}, where subsequent numerical computations are based on the fit results. 

Since in that particular case, the main idea was to capture and quantitatively include residual contributions from poles inside the integration domain, the precise location of the poles might actually be of minor importance.
In other possible circumstances, however, where the precise positions do matter, one must take care to ensure a more faithful representation.
It is for cases like these that the approach proposed in the present paper will prove advantageous, since our classifiers or others like it help to pre-filter and better direct the regressional computational effort.

\paragraph{Real-world limitations}
After these concrete results from our approach, we would like to delineate a few areas that point to the limits of this and standard approaches.

One such limitation regards the capacity of the method to deal with many additional poles, regardless where they are.
Considering that any fitting model that allows for more than one pole (pair) for each pole essentially introduces more and more free parameters, this can easily de-stabilize the fit.
The same is true for our approach, e.\,g., when the classifier tends to favor multi-pole scenarios over single-pole ones, or  when we encounter convergence issues.

Other issues can be caused by numerical artifacts.
The prime example in our particular use case come from divisions of pairs of very small numbers close to zero on the numerical integration grid.
While a thorough solution will try to avoid such artifacts completely, this is not always easily possible in applications close to physical and other use cases.

Finally, it is noteworthy that also standard approaches fail at describing real-world data.
As a consequence, encountering some difficulties in our approach does not necessarily correspond to a tradeoff in terms of getting advantages but at the same time encountering those difficulties.
Actually, the advantages may as well be counted as a straight-up improvement.

\section{Conclusion and outlook}\label{conclusion}

Based on a use case from theoretical hadron physics, we have detailed a set of methods for analyzing the configuration and locations of the dominant singularities in the complex plane for a function known only on a discrete integration grid on the positive real axis.

Naturally, our results for our starting use case are novel and interesting on an immediate level.
They are relevant for the study of confinement, but also for other concrete problems in the realm of QCD Green functions.
In particular, in the context of the nonperturbative framework provided by the simultaneous analysis of coupled DSEs and BSEs, more reliable information on dominant singularities can be enormously beneficial.

In essence, however, this problem is a general one.
As a consequence, our setup could help to improve upon the numerical path to similar solutions in any general kind of situation where it is important to know the configuration of the dominant singularity in a physical problem.

Still more generally, our ways to combine conventional techniques with ANN approaches at different stages in the solution procedure are a typical (and successful) example of the fruitful interconnection of modern machine-learning with traditional numerical approaches.

\appendix
\section{Technicalities}\label{technicalities}
In order to be able to talk about model details in a truncated version of the quark gap equation, it is necessary to provide at least some details on the model structure of the quark self-energy. 
To do this, we rewrite the self-energy term from Eq.~(\ref{eq:dse}), which reads
\begin{equation}\label{eq:dse_detail}
  \Sigma(p) = \int_q  D^{\mu\nu}(k)\, \gamma^\mu \,S(q)\,  \Gamma^\nu(q,p)\,  .
\end{equation}
While the structure of the bare quark-gluon vertex $\gamma^\mu$ is explicit and the structure of the dressed quark propagator $S(q)$ is known (see Eq.~(\ref{eq:dressedpropagator})), the other two parts of this expression are dressed quantities that satisfy their own DSEs.

From theory, we know the possible covariant basis structures for these objects and would be able to draw conclusions about the associated dressing functions from the relevant DSEs.
Furthermore, constraints are inferable via identities among such objects that stem from various other aspects of the theory, like conservation laws, for example.
Gathering information from the DSEs and using the interrelations of their solutions leads to an infinite tower of coupled and, in general, nonlinear integral equations.

In practice, this entails that for a numerical solution some kind of truncation must be made, and this step determines the form of the dressed gluon propagator $D^{\mu\nu}(k)$ and the dressed quark-gluon vertex $\Gamma^\nu(q,p)$.
Without much explanation, which can be found elsewhere (e.g., \cite{Maris:2003vk}), we simply write down the ladder-approximation of the quark self-energy here for easy reference (e.g., for understanding the code in the associated Github repository).
Effectively, the product of dressed gluon propagator and dressed quark-gluon vertex is replaced by a bare-vertex times bare-propagator structure multiplied by relevant factors and an effective interaction term, and reads:
\begin{equation}\label{eq:dse_detail}
  \Sigma_{RL}(p) = \frac{4}{3}\int_q  \mathcal{G}(k^2)\,D_f^{\mu\nu}(k)\, \gamma^\mu \,S(q)\,  \gamma^\nu\,  .
\end{equation}

Here, $D_f^{\mu\nu}(k)$ is the free gluon propagator, $\gamma^\nu$ is the bare quark-gluon vertex, $\mathcal{G}(k^2)$ is the aforementioned effective interaction, and the factor $\frac{4}{3}$ comes from the color trace. 
In this way, the model and its parameters are contained in the function $\mathcal{G}(k^2)$, which is parameterized as \cite{Maris:1999nt}
\begin{eqnarray}\nonumber
\frac{{\cal G}(k^2)}{k^2}& =& \frac{4\pi^2 D}{\omega^6} k^2\;\mathrm{e}^{-k^2/\omega^2}\\\label{eq:effective_interaction}
&+&\frac{4\pi\;\gamma_m \pi\;\mathcal{F}(k^2) }{1/2 \ln [\tau\!+\!(1\!+\!k^2/\Lambda_\mathrm{QCD}^2)^2]}.
\end{eqnarray}
In order to explain the origin and meaning of these parameters, we would like to make a few remarks, knowing well that a satisfactory explanation is beyond the scope of the current presentation. 
Thus, we motivate each parameter only briefly and refer the interested reader to the relevant references.

The functional form itself comes from both the desire to work phenomenologically as well as the need to stay in line with theoretical requirements.
For example, this parameterization (MT, \cite{Maris:1999nt}) reproduces the correct perturbative limit for large momentum-squared. 
More precisely, it preserves the one-loop renormalization-group behavior of QCD for solutions of the quark DSE. 
There, certain parameters and functional forms in addition were geared towards numerical tractability, a point relevant also for the present article. 
In particular, we have ${\cal F}(k^2)= [1 - \exp(-k^2/[4 m_t^2])]/k^2$, with the parameter values $m_t=0.5$~GeV (an intermediate mass scale for the exponential used in lifting one of the IR singularities), $\tau={\rm e}^2-1$ (an intermediate scale to help modify the analytic structure of the log term in the perturbative interaction), $N_f=4$ (the number of active flavors in the UV), $\Lambda_\mathrm{QCD}^{N_f=4}= 0.234\,{\rm GeV}$ (a phenomenologycally motivated value for an important mass scale in QCD), and $\gamma_m=12/(33-2N_f)$ (the corresponding value for the anomalous dimension)  \cite{Maris:1999nt}.
Part of these parameterizations' design is also the coupling's behavior for small momentum squared (in the IR). 
Phenomenologically, the impact of the effective coupling's far IR behavior, e.\,g., on meson masses, is expected to be small \cite{Blank:2010pa}, while certain functional forms facilitate numerical treatment.

In another version of the interaction (AWW, \cite{Alkofer:2002bp}), one drops the one-loop UV contribution completely, which also remedies a few of the numerical problems, in particular those related to the log term and its rather complicated analytic structure. 
Both kinds of interactions have been used in the literature for the phenomenological study of hadrons and their properties.
Herein, we combine several parameter sets from each form in our analysis for both inter- and intra-model comparison.

The remaining parameters, $D$, in GeV${}^2$ and $\omega$, in GeV, are indicative of the long-range part of the effective interaction in that they determine how high ($D$) and wide ($\omega$) in momentum space the interaction curve is.
The product of these two parameters has been found to represent a coupling strength that phenomenologically corresponds to a certain value of the quark condensate and can therefore be fit accordingly.

Finally, the current-quark mass introduced via the mass term (the scalar part of $S_0^{-1}(p)$) in the quark DSE (\ref{eq:dse}) completes the set of open parameters in the model.
Its value lies in the typical range of suitable masses, and to make things simpler, we refer to two different mass ranges as \emph{light} and \emph{strange} in this context.
These two cases are the most important in our study, since they correspond to mass poles closest to the origin in the quark propagator's dressing functions. 
For higher flavors, also these poles move further out and thus do not relate well to the particular techniques we have laid out herein.

This phenomenological setup and the resulting set of parameters is what leads to the compilation of parameters in Tab.~\ref{tablemodelparams}, which represents a collection of typical parameter combinations which can serve as a suitable testing ground for our methods. 
In addition to the parameters, we also introduce a vertex structure more complex than the bare vertex. 
While the summary of parameters used with each vertex structure is included in Tab.~\ref{tablemodelparams}, and the table is enough for comprehensive reference in this analysis, we present the other vertex structure in \ref{bcv}.

\begin{table}[!ht]
	\caption{Various models and sets of model parameters in combination with two vertex structures. 
	Model parameters correspond to the functional form of the effective coupling as given in the text.
	Vertex structures refer to the two distinct cases described in the text.
	The model names are used in the main part of the paper to not overwhelm the reader with unnecessary technical information.
	This table serves as the reference to decoding these model names as well as for the interested reader to get a quick overview of the parameter used in terms of the effective interaction.}				\label{tablemodelparams}
	\begin{center}
		\begin{tabularx}{\columnwidth}{lccrrc}\hline
			Name & Vertex & Model  & $\omega$  & $D$ & $m_q$  \\ \hline
			M--BV--1 & bare & MT & 0.3 & 1.24 & light \\
			M--BV--2 & bare & MT & 0.4 & 0.93 & light \\
			M--BV--3 & bare & MT & 0.5 & 0.744 & light \\
			A--BV--1 & bare & AWW & 0.4 & 1.152 & strange \\
			A--BV--2 & bare & AWW & 0.55 & 0.84 & light \\
			M--BC--1 & BC & MT & 0.3 & 0.1 & light \\
			M--BC--2 & BC & MT & 0.4 & 0.1 & light \\
			M--BC--3 & BC & MT & 0.5 & 0.1 & light \\
			A--BC--1 & BC & AWW & 0.4 & 0.1 & strange \\
			A--BC--2 & BC & AWW & 0.55 & 0.1 & light \\ \hline
		\end{tabularx}
	\end{center}
\end{table}

\section{The Ball-Chiu vertex ansatz}\label{bcv}
In this appendix, we briefly sketch the structure and the relevant details of the Ball-Chiu ansatz for the dressed quark-gluon vertex.
The motivation for this ansatz as well as further improvements are explained in detail in Ref.~\cite{Curtis:1990zs}.
The main point is the vertex ansatz's satisfaction of the relevant Ward-Takahashi identities plus the corresponding Ward identity in the limit of $(p-q)\rightarrow 0$, which leads to the following structure:
\begin{eqnarray}\nonumber
 \!\!\Gamma^\mu(q,p)&=&\frac{1}{2}\,\gamma^\mu [A(q^2)+A(p^2)]\\\nonumber
 &+&\frac{1}{2}\,(q^\mu+p^\mu) (\gamma\!\cdot\! q+\gamma\!\cdot\! p) \frac{[A(q^2)-A(p^2)]}{q^2-p^2}\\
 &+&(q^\mu+p^\mu)  \frac{[B(q^2)-B(p^2)]}{q^2-p^2}\label{eq:bcv}
\end{eqnarray}
All terms involved here have already been defined previously, which makes this construction an interesting candidate for a non-trivial vertex.

\section{Visualizations of class-predictor results}\label{app:classpredictions}
\begin{figure}
	\centering
    \subfloat[Class 0]{%
        \includegraphics[width=0.19\textwidth]{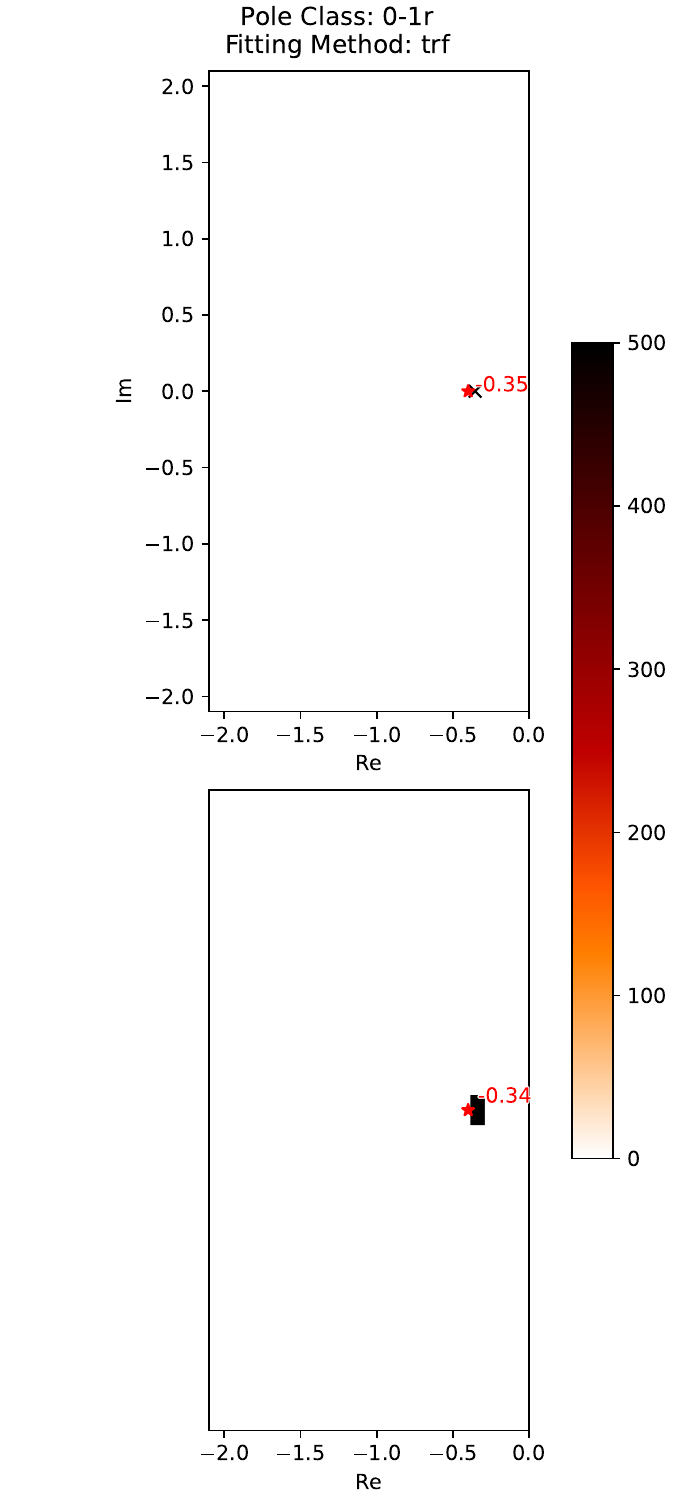}
        \label{fig:rsub1}
    }
    \hfill
    \subfloat[Class 1]{%
        \includegraphics[width=0.19\textwidth]{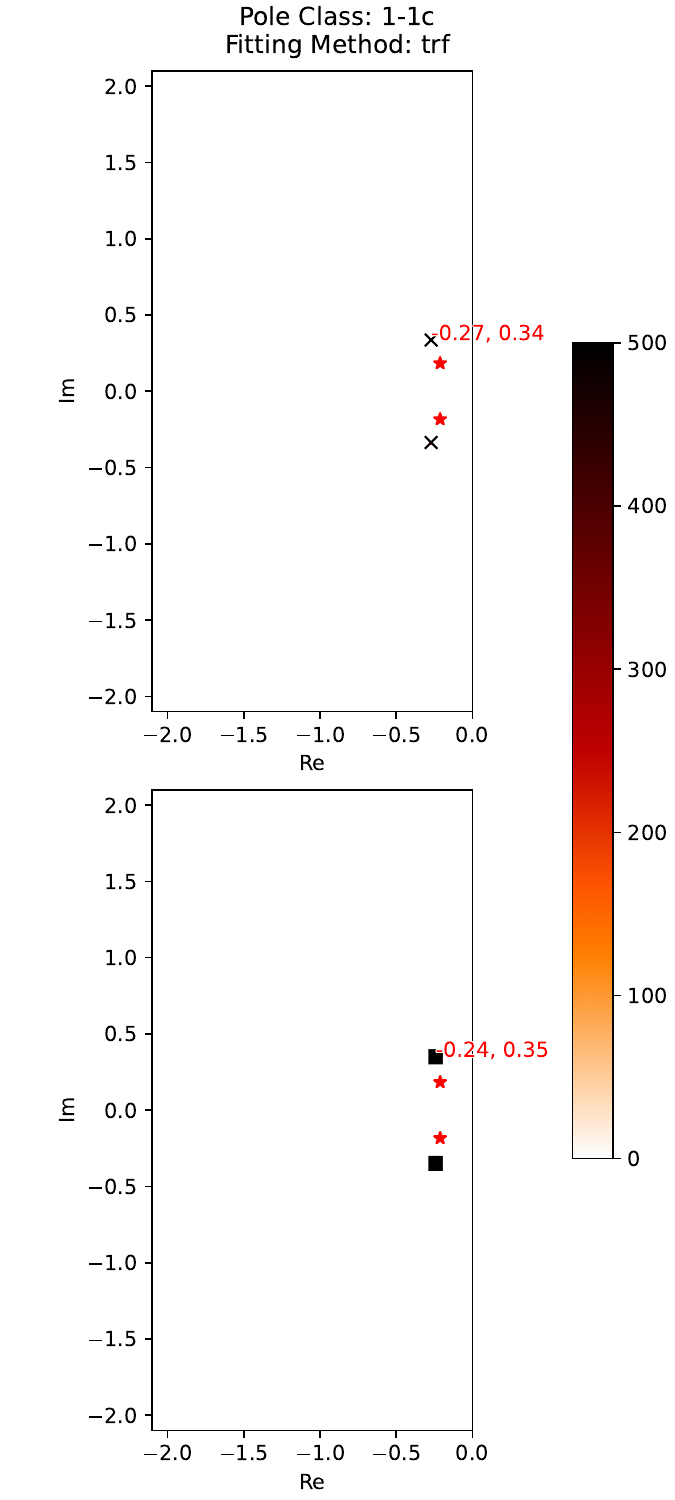}
        \label{fig:rsub2}
    }
	  \caption{Visualizations of the regressor- and TRF-predictions for classes $0$ and $1$ for model M--BV--2. For a description, see text in Sec.~\ref{application}.
	  \label{fig:appc1}}
\end{figure}

\begin{figure*}
	\centering
    \subfloat[Class 2]{%
        \includegraphics[width=0.3\textwidth]{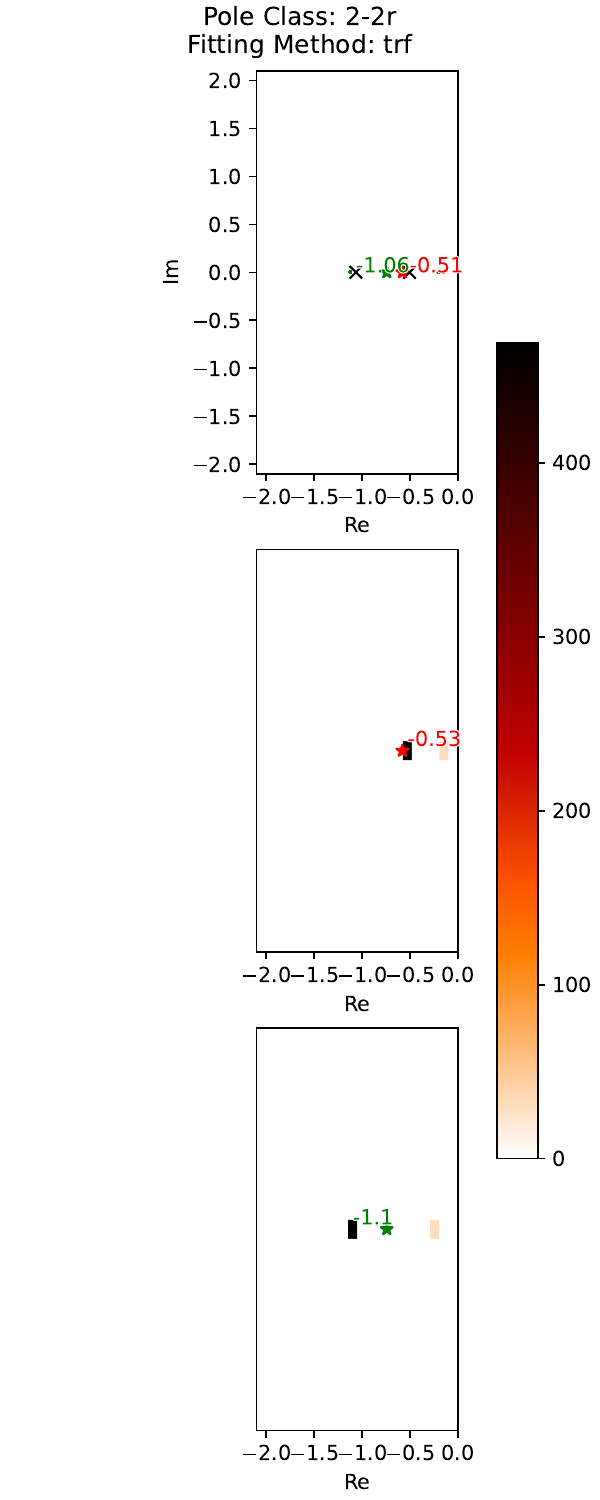}
        \label{fig:rsub3}
    }\hfill
    \subfloat[Class 3]{%
        \includegraphics[width=0.3\textwidth]{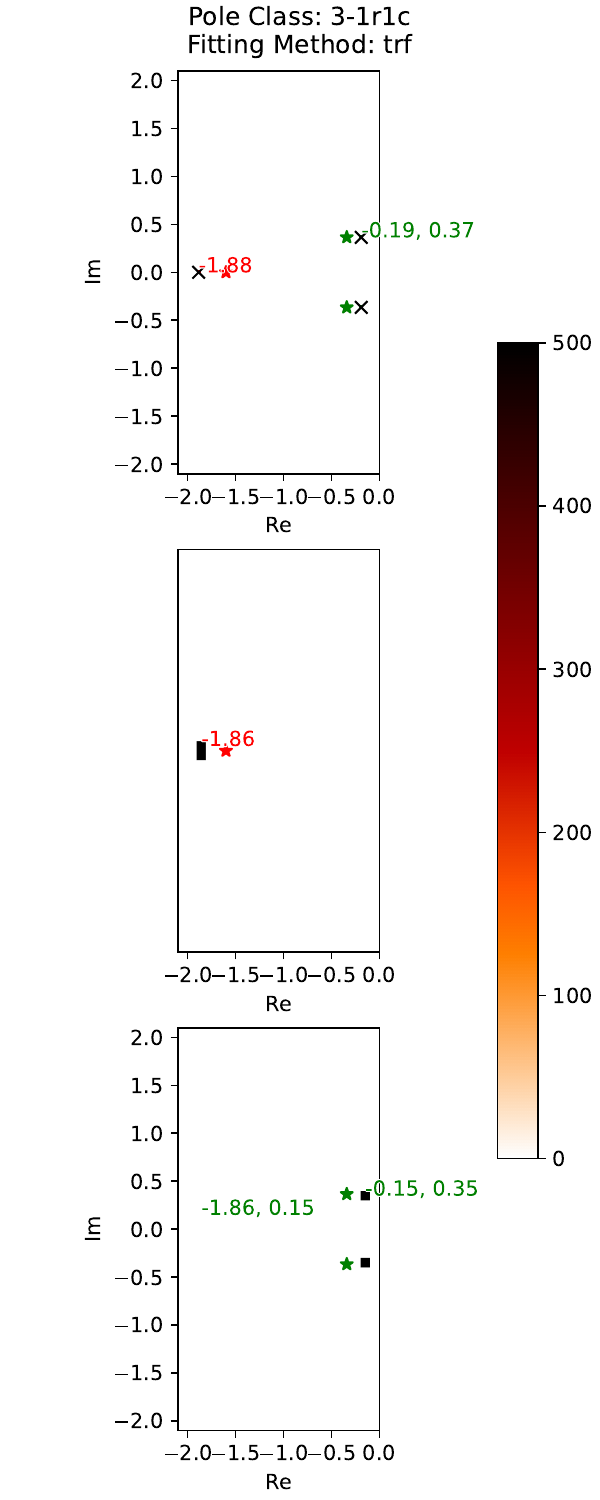}
        \label{fig:rsub4}
    }
    \hfill    
    \subfloat[Class 4]{%
        \includegraphics[width=0.3\textwidth]{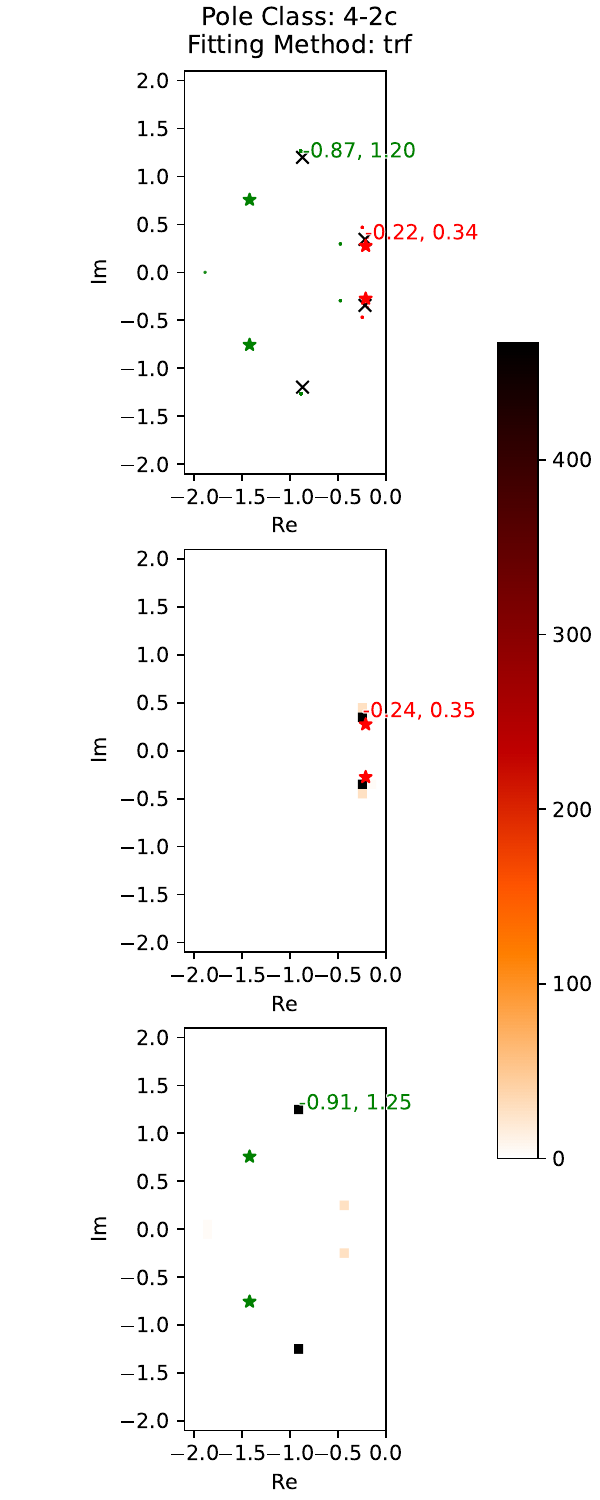}
        \label{fig:rsub5}
    }	  \caption{Visualizations of the regressor- and TRF-predictions for classes $2$, $3$, and $4$ for model M--BV--2. For a description, see text in Sec.~\ref{application}.
	  \label{fig:appc2}}
\end{figure*}

\begin{figure*}
	\centering
    \subfloat[Class 5]{%
        \includegraphics[width=0.45\textwidth]{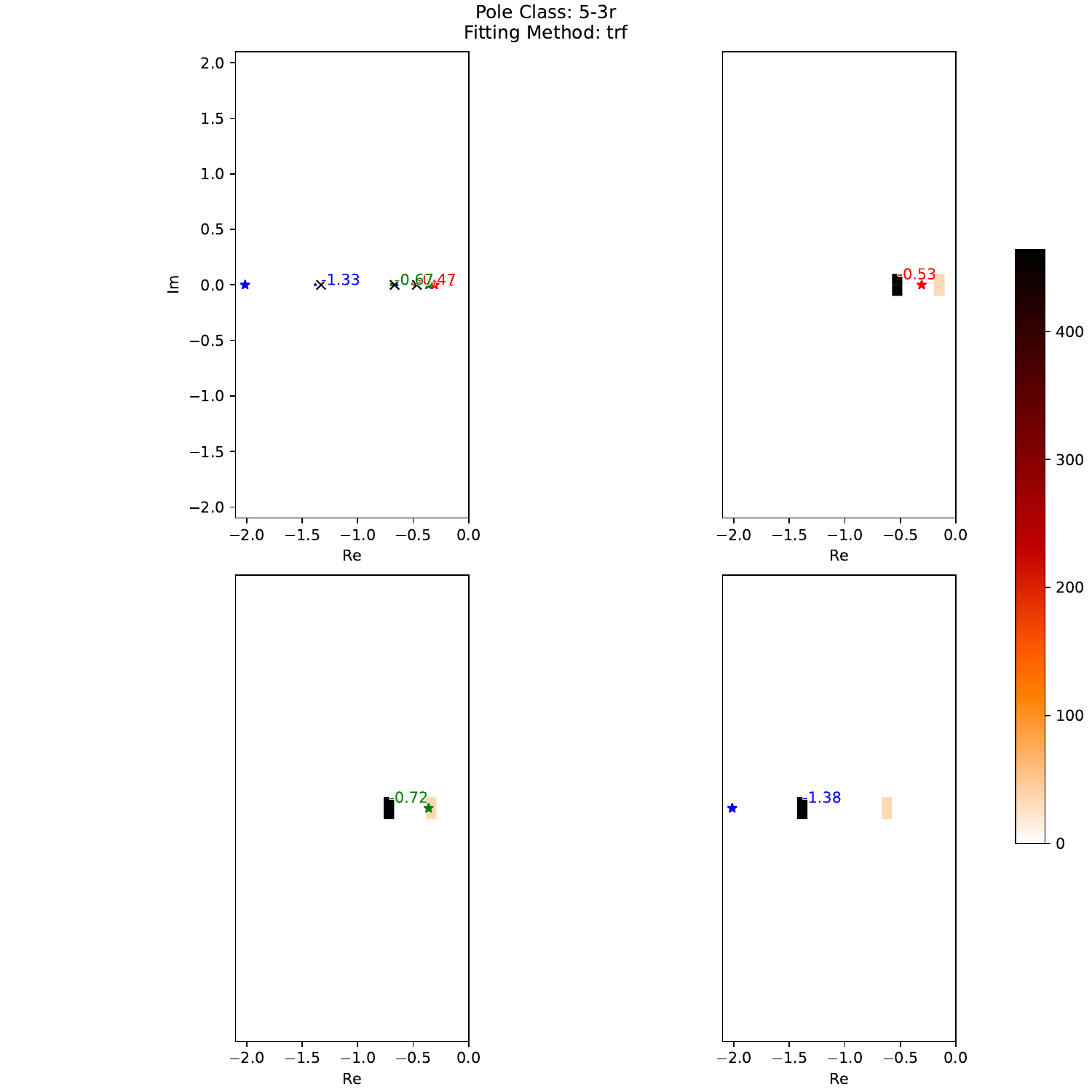}
        \label{fig:rsub6}
    }  
    \hfill    
    \subfloat[Class 6]{%
        \includegraphics[width=0.45\textwidth]{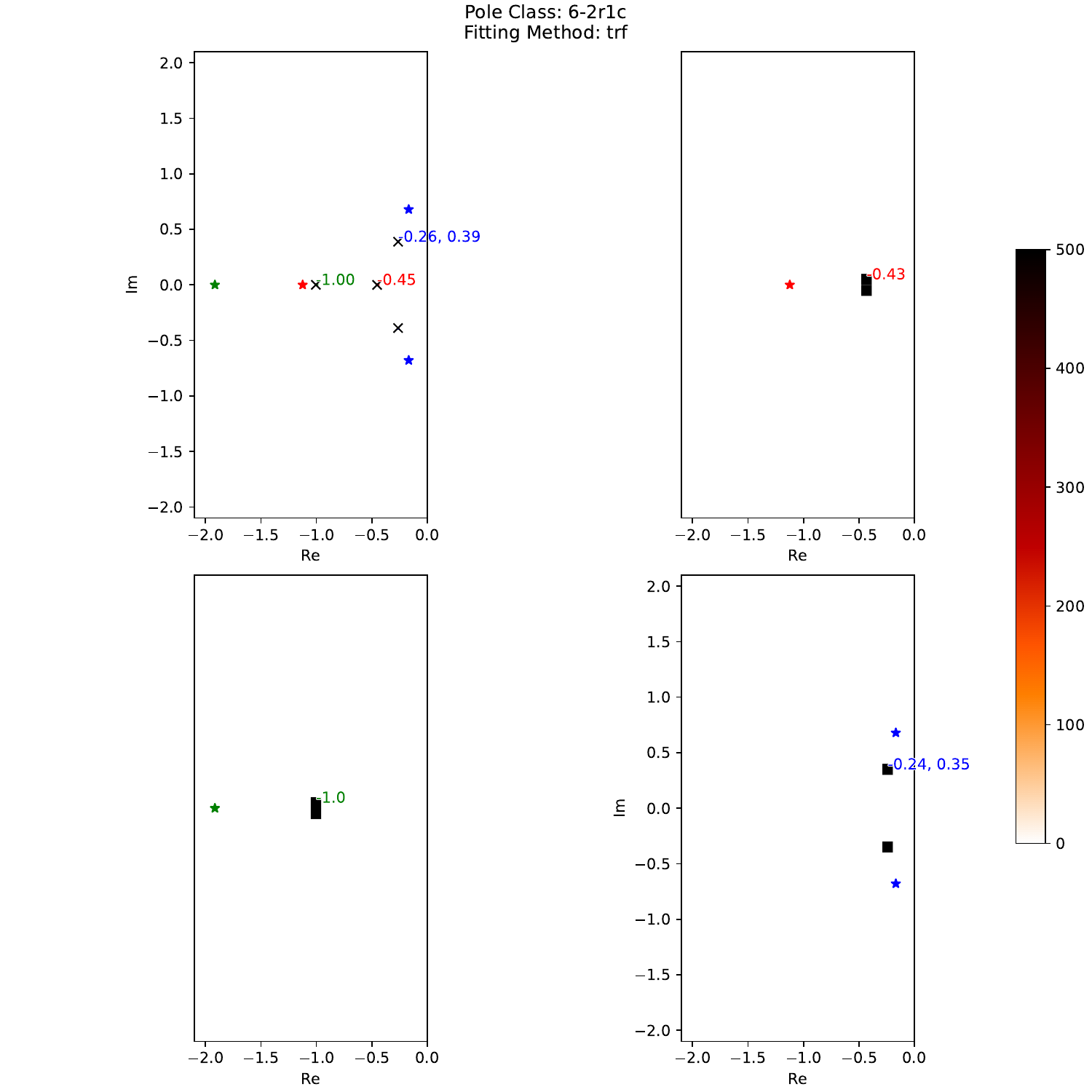}
        \label{fig:rsub7}
    }\\     
    \subfloat[Class 7]{%
        \includegraphics[width=0.45\textwidth]{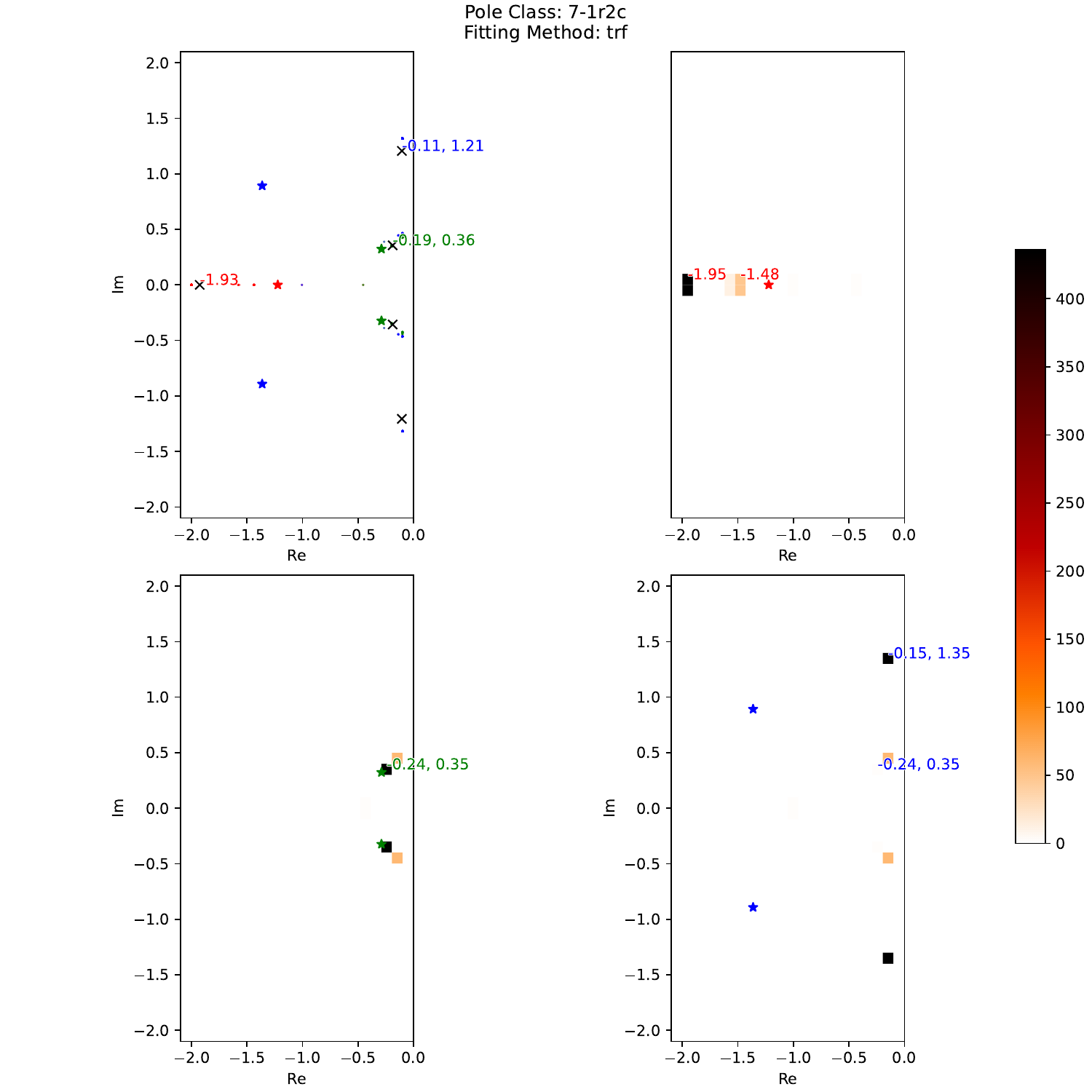}
        \label{fig:rsub8}
    }
    \hfill    
    \subfloat[Class 8]{%
        \includegraphics[width=0.45\textwidth]{figs/M-BV-2_class_8.pdf}
        \label{fig:rsub9}
    }
	  \caption{Visualizations of the regressor- and TRF-predictions for classes $5$, $6$, $7$, and $8$ for model M--BV--2. For a description, see text in Sec.~\ref{application}.
	  \label{fig:appc3}}
\end{figure*}

In this appendix, we present a complete set of visualizations of the regressor- and TRF-predictions made for the model M--BV--2, as exemplified in Sec.~\ref{application}. 
For a detailed description of the elements in each subfigure, please consult the text there. 
Also note that these figures correspond to the numbers as given in Tab.~\ref{results:regressor}.
Fig.~\ref{fig:appc1} shows the visualizations for classes $0$ and $1$, Fig.~\ref{fig:appc2} shows the visualizations for classes $2$, $3$, and $4$, and Fig.~\ref{fig:appc3} shows the visualizations for classes $5$, $6$, $7$, and $8$, 

\section*{Acknowledgements} The work of S.\,K., A.\,K., and W.\,L. was supported in part by the Austrian Academy of Sciences Innovation Fund Project "Machine Learning for Theoretical Particle Physics (ML-TPP)". The work of T.H. did not receive financial support from Carl Zeiss SMT GmbH.

\end{document}